\documentclass{ptephy_v1}

\usepackage{amsmath,amssymb}
\usepackage{subfig}
\usepackage{url}
\usepackage{tikz}
\usepackage{float}
\usepackage{braket}

\newcommand{\bm}[1]{\mbox{\boldmath $#1$}}
\newcommand{\OD}[1]{\frac{\mathrm{d}}{\mathrm{d} #1}}

\newcommand{\ODD}[2]{\frac{\mathrm{d} #1}{\mathrm{d} #2}}

\renewcommand{\Re}[1]{\mathrm{Re}\left[ #1 \right]}
\renewcommand{\Im}[1]{\mathrm{Im}\left[ #1 \right]}

\begin{document}

\title{Adiabatic evolution of the self-interacting axion field around rotating black holes}


\author[1,*]{Hidetoshi Omiya}
\affil{Department of Physics, Kyoto University, Kyoto 606-8502, Japan}

\author[1]{Takuya Takahashi}

\author[1,2]{Takahiro Tanaka}
\affil{Center for Gravitational Physics, Yukawa Institute for Theoretical Physics, Kyoto University, Kyoto 606-8502, Japan}

\email{omiya@tap.scphys.kyoto-u.ac.jp}

\begin{abstract}%
Ultra light axion fields, motivated by the string theory, form a large condensate (axion cloud) around rotating black holes through superradiant instability. Several effects due to the axion cloud, such as the spin-down of black holes and the emission of monochromatic gravitational waves, open a new window to search for axions by astrophysical observations. When the axion self-interaction is considered, the evolution of cloud is altered significantly, and an explosive phenomenon called bosenova can happen. Thus, it is necessary to understand the precise evolution of  self-interacting clouds for the detection of axions by astrophysical observations.
In this paper, we propose a new method to track the whole process of the growth of self-interacting axion clouds employing the adiabatic approximation. We emphasize that our method relies neither on the non-relativistic approximation nor on perturbative treatment of the self-interaction, which is often used in literature. 
Our main finding is that the evolution of cloud in the strongly self-interacting regime depends on the strength of the gravitational coupling between the axion and the black hole. For a large coupling, the cloud evolves into a quasi-stationary state where the superradiant energy gain is balanced with the energy dissipation to infinity by the self-interaction. On the other hand, when one decreases the size of coupling, clouds become unstable at some energy, which would be interpreted as the onset of bosenova.
\end{abstract}

\maketitle

\section{Introduction}

Axion forms a class of the most motivated particles beyond the standard model. It has the potential to solve the strong CP problem \cite{Peccei:1977hh, Weinberg:1977ma,Wilczek:1977pj,Kim:1979if,Shifman:1979if,Zhitnitsky:1980tq,Dine:1981rt} and can be a candidate of dark matter \cite{Dine:1982ah,Preskill:1982cy,Abbott:1982af,Hui:2016ltb}. 
In addition, they can be naturally derived from sting theory \cite{Svrcek:2006yi}. 
An interesting point is that string theory predicts the plenitude of axions in our Universe and the Compton wavelength of axions can be comparable to the astrophysical scale. 
This opens the possibility of observing axions through astrophysical phenomena \cite{Arvanitaki:2009fg}. 
In this paper, we focus on the phenomena related to black hole (BH) physics.

Let us consider an axion field around a spinning BH. 
Since the axion has small but non-zero mass $\mu$, they are bounded by the gravitational potential of the BH. 
At the same time, axion extracts the energy and angular momentum from the BH by the superadiance (see \cite{Brito:2015oca} for the detail of superradiance). 
This indicates the existence of an instability, known as {\it superradiant instability} \cite{Zouros:1979iw,Detweiler:1980uk,Dolan:2007mj}. 
The time scale of instability can be much smaller than the age of the Universe 
when the Compton wavelength of the axion is comparable to the size of the BH. 
Thus, the axion with a mass comparable to the astrophysical scale forms a large condensate around the BH 
by the superradiant instability. 
We refer to this condensate as an {\it axion cloud} in this paper.

The axion cloud will induce several phenomena which can be observed \cite{Arvanitaki:2010sy}. 
One is the spin-down of the BH, owing to the angular momentum extraction by the axion cloud. 
Thus, the presence of an axion with a corresponding mass excludes highly spinning BHs and predicts a characteristic distribution of the BH mass and spin \cite{Brito:2014wla,Stott:2018opm,Fernandez:2019qbj,Ng:2019jsx,Ng:2020ruv}. 
Other phenomena are the emission of characteristic gravitational waves from the cloud associated with 
the level transition similar to the photon emission in the hydrogen atom or the pair annihilation of axions \cite{Yoshino:2013ofa, Arvanitaki:2014wva, Zhu:2020tht, Tsukada:2018mbp, LIGOScientific:2021jlr} 
as well as the modification of the gravitational wave form from binary BHs \cite{Baumann:2018vus,  Ding:2020bnl, Choudhary:2020pxy,Su:2021dwz,Takahashi:2021yhy, Baumann:2021fkf}.

If one includes the self-interaction of the axion more dramatic phenomena can happen. 
In general, axion has a nonlinear potential induced by non-perturbative quantum effects 
and the leading order interaction is typically attractive. 
Therefore, when a cloud grows to a large amplitude, the attractive force due to the self-interaction might 
induce a collapse of the cloud. This collapse is called {\it bosenova} and the burst of gravitational waves 
is expected to be generated during the collapse \cite{Arvanitaki:2010sy, Yoshino:2012kn, Yoshino:2015nsa}. 
Besides the bosenova, the self-interaction can cause the energy loss of the cloud through several channels \cite{Arvanitaki:2010sy,Baryakhtar:2020gao,Omiya:2020vji}. 
These effects have potential of terminating the superradiant instability and prevents the occurrence of the bosenova.

For the future detection of axion clouds through observations, one must know the precise evolution of the cloud including the self-interaction. In literature only few works take into account the effect of self-interaction extensively. 
One is the work on numerical simulations \cite{Yoshino:2012kn, Yoshino:2015nsa}, 
which suggests the occurrence of bosenova. 
However, the previous dynamical simulations are not satisfactory due to  
the ambiguity in the choice of the initial condition.
Because of the large discrepancy between the dynamical time scale and instability time scale, 
it is hard to perform a long-term simulation starting with a small amplitude of the cloud where the linear approximation is valid. For this reason, it was necessary to give a naive guess about the initial condition for the numerical simulation, which is, in fact, simply given by scaling the solution of the linearized equation in the previous works. 
Since the cloud starts with a very small amplitude\footnote{If we assume cloud started with an amplitude around mass of axion, then the mass of the cloud is around $\sim 10^{-76}M/(G\mu M)^4 (\mu/10^{-10} {\rm eV})^2$.} and change the shape by the effect of self-interaction as it grows, it is difficult to justify the usage of a linearized solution with a large amplitude as the initial condition to simulate a realistic situation. 

Other works \cite{Arvanitaki:2010sy,Baryakhtar:2020gao,Omiya:2020vji} treat the self-interaction perturbatively, and often adopt the non-relativistic approximation ($G \mu M \ll 1$, $M$ is BH mass and $c=\hbar=1$). 
When the self-interaction becomes important, the perturbative treatment breaks down \cite{Omiya:2020vji}. 
Therefore, the evolution of cloud in the strongly self-interacting regime, where bosenova may occur, cannot be investigated using perturbation theory. 
Furthermore, the non-relativistic approximation cannot treat the most interesting parameter region 
where the instability time scale is maximized ($G \mu M \sim 1$). To summarize, both dynamical simulations and perturbative treatment are not satisfactory.

To overcome this situation, we develop a new method to track the evolution of clouds starting with a single superradiant mode, without relying on either perturbative or non-relativistic treatment. Our basic strategy is to use the fact that the evolution of cloud is very slow compared to the dynamical time scale, even if the perturbation theory breaks down \cite{Omiya:2020vji}. 
Then, one can approximate the cloud by a stationary configuration with a given amplitude within the time scale much shorter than that of the superradiant instability. After obtaining a sequence of solutions parametrized by the cloud amplitude, the time evolution of the amplitude is determined by the energy balance argument. 
Our method reveals that the final fate of the cloud basically depends on the strength of the gravitational coupling between the axion and the BH, $G\mu M$. For a large coupling ($G\mu M \gtrsim 0.32$), 
the final state of the cloud becomes quasi-stationary, where the energy gain by the superradiance is balanced with the energy dissipation to infinity induced by the self-interaction. 
For a small coupling ($G\mu M \lesssim 0.32$), the onset of the dynamical instability is suggested. 
This instability can be expected to lead to the ignition of a bosenova. 
In addition, our calculation gives the deformation of the cloud by the self-interaction, which turned out to be  significantly large.

This paper is organized as follows. In section \ref{sec:2}, we review the superradiant instability of axion around rotating BHs. In section \ref{sec:3}, we present a method to track the adiabatic evolution of a self-interacting axion field around a rotating BH. In section \ref{sec:4} we show the result of numerical calculations. 
In section \ref{sec:5}, we present a toy model of the axion cloud, which explains the behavior of the cloud numerically obtained in section \ref{sec:4}. 
In section \ref{sec:6}, we summarize our results and briefly comment on the effects we have not taken into account. 
In the following of this paper, we take units $c= G = \hbar = 1$, unless otherwise stated.

\section{Axion cloud around a black hole}\label{sec:2}

In this section, we review how axion clouds are formed around Kerr black holes by the superradiant instability. 
For further details on the superradiant instability, see Ref. \cite{Brito:2015oca}. 

In the rest of this paper, we consider an axion field $\phi$ whose action is given by  
\begin{align}\label{eq:action}
	S = F_a^2 \int d^4\! x \sqrt{-g} \left\{ - \frac{1}{2}g^{\mu\nu}\partial_\mu \phi \partial_\nu \phi - V(\phi)\right\}~,
\end{align}
where $V(\phi)$ is the potential of the axion induced by some quantum effects. 
In this paper, we employ 
\begin{align}\label{eq:cospotential}
	V(\phi) = \mu^2 \left(1 - \cos\phi\right)~,
\end{align}
which is the well-known potential for the QCD axion \cite{Weinberg:1996kr}. 
Note that we normalize $\phi$ by the decay constant $F_a$ to make $\phi$ non-dimensional. 
Here, $g_{\mu\nu}$ is the Kerr metric specified by
\begin{align}
\label{kerrmetric}
	ds^2 &=  - \left(1 - \frac{2 M r}{\rho^2}\right)dt^2 - \frac{4 a M r\sin^2\theta}{\rho^2} dt\, d\varphi \cr
	 &\ \ \ \ \ + \left[(r^2 + a^2)\sin^2\theta+ \frac{2 M r}{\rho^2} a^2 \sin^4 \theta \right]d\varphi^2 + \frac{\rho^2}{\Delta} dr^2 + \rho^2 d\theta^2~,
\end{align}
with 
\begin{align}
	\Delta &= r^2 - 2 Mr + a^2~, & \rho^2 &= r^2 + a^2\cos^2\theta~,
\end{align}
in the Boyer-Lindquist (BL) coordinates. 
Solutions to $\Delta = 0$ give the location of the event horizon $r_+ = M + \sqrt{M^2 - a^2}$ and that of the Cauchy horizon $r_- = M - \sqrt{M^2 - a^2}$.  
For the action \eqref{eq:action}, the equation of motion for the axion field is
\begin{align}
	\square_g \phi - V'(\phi) = 0~,
	\label{eq:eomfull}
\end{align}
where $\square_g$ is the d'Alembertian on the Kerr metric.

When the amplitude of the axion field is small, we can ignore the higher order terms in the axion potential \eqref{eq:cospotential} and thus approximate the potential as
\begin{align}\label{eq:potlin}
	V(\phi) \sim \frac{1}{2}\mu^2 \phi^2~.
\end{align}
Then, the equation of motion takes a linearized form
\begin{align}\label{eq:eomlin}
	\left(\square_g - \mu^2\right)\phi = 0~,
\end{align}
which can be solved by the separation of variables \cite{Brill:1972xj}. 
Taking the ansatz
\begin{align}
	\label{eq:sep}
	\phi = \Re{e^{-i(\omega t - m \varphi)}S_{lm\omega}(\theta)R_{lm\omega}(r)}~,
\end{align}
Eq. \eqref{eq:eomlin} is decomposed into two ordinary differential equations
\begin{gather}
		\label{eq:angularEOM}
		\frac{1}{\sin\theta}\OD{\theta}\left(\sin\theta \ODD{{S}_{lm\omega}}{\theta}\right) + \left[c^2(\omega) \cos^2\theta - \frac{m^2}{\sin^2\theta}\right]S_{lm\omega} = - \Lambda_{lm}(\omega) S_{lm\omega}~,
\end{gather}
and
\begin{gather}
		\label{eq:radialEOM}
		\OD{r}\left(\Delta\ODD{R_{lm\omega}}{r}\right)  + \left[\frac{K^2(\omega)}{\Delta} - \mu^2 r^2 -\lambda_{lm}(\omega) \right]R_{lm\omega} = 0~,
\end{gather}
where
\begin{align}
		 c^2 (\omega) &= a^2 (\omega^2 - \mu^2)~, \qquad K(\omega) = (r^2+a^2)\omega - am~,\cr
	 \lambda_{lm}(\omega) & = -2am \omega +a^2\omega^2 +\Lambda_{lm}(\omega)~.
\end{align}

\begin{figure}[t]
        \centering
	\includegraphics[keepaspectratio,scale=0.3]{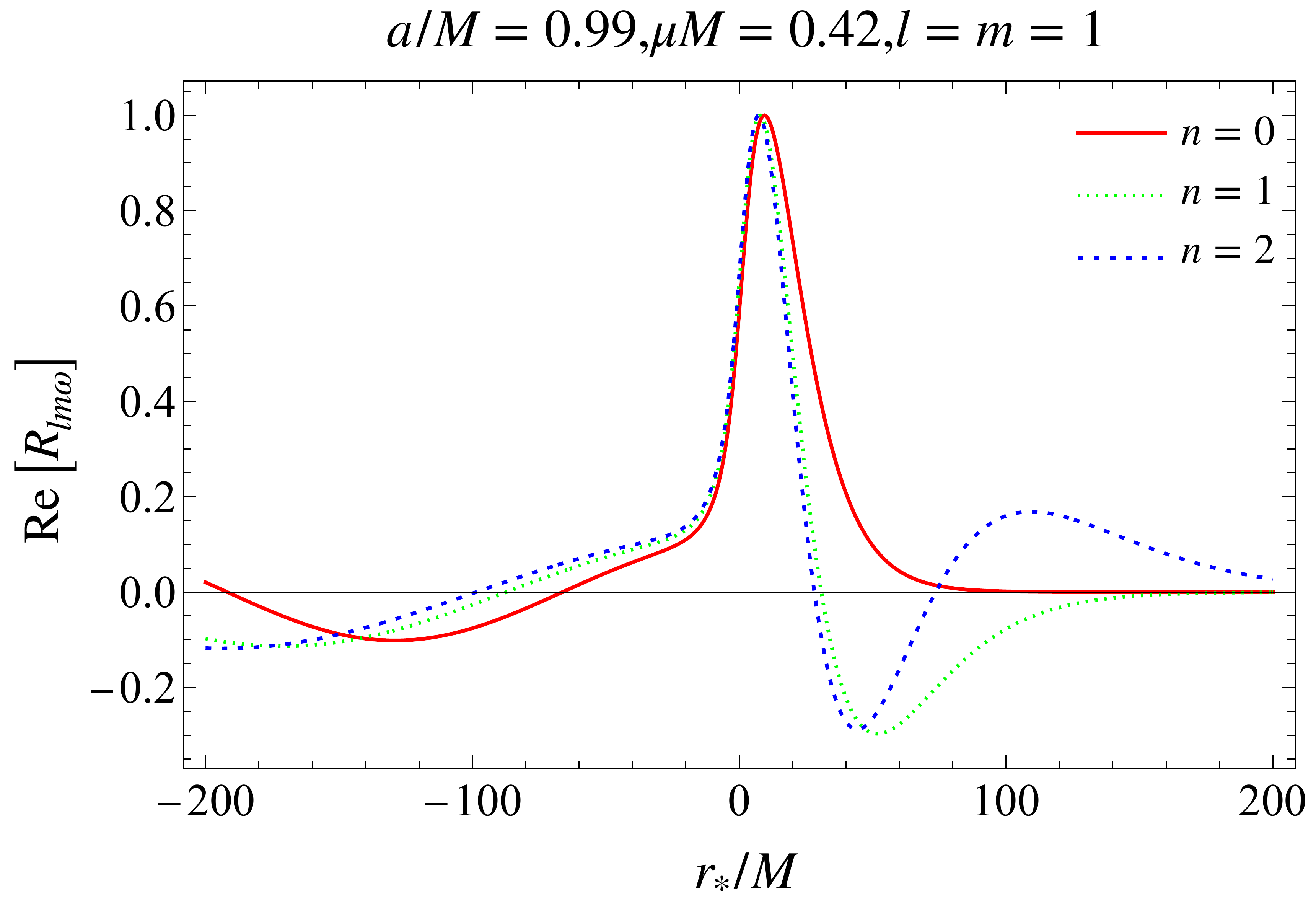}
	\caption{The real part of the radial mode function $R_{lm\omega}$ with principal quantum number $n=0,1,2$, and $l=m=1$. Mode functions are normalized for their peak amplitude to be unity. We take the spin of the central black hole as $a/M =0.99$ and the mass of the axion as $\mu M = 0.42$, which gives the maximum growth rate. Here, $r_*$ is defined as $dr_* = (r^2 + a^2)dr/\Delta$.}
	\label{fig:linconfig}
\end{figure}
The solution to Eq. \eqref{eq:radialEOM} with the ingoing boundary condition at the event horizon and exponential fall off at infinity is very similar to the wave function of the hydrogen atom \cite{Detweiler:1980uk, Dolan:2007mj} (see Fig. \ref{fig:linconfig} for the configuration). Solutions are labeled by $(\omega,l,m)$, and $\omega$ takes discrete values, labeled by $n$, as in the case of the energy levels of hydrogen atom.
When the cloud is less massive, the initial frequency $\omega$ is supposed to satisfy 
\begin{align}
	\omega_R &< \mu~,\\
	\label{eq:SRcond}
	\frac{|\omega|^2}{\omega_R} <& m \Omega_H~,\\
	\omega_I &> 0~,
\end{align}
with $\omega_R = \Re{\omega}, \omega_I = \Im{\omega}$,and $\Omega_H = a/2M r_+$. 
The third condition implies the presence of instability, which is expected when the first and second conditions are satisfied. The first condition means that axions are bounded by the gravitational potential, while the second condition is the superradiance condition, which states that the axion field is extracting the energy and the angular momentum from the BH. 
Thus, the trapped axions around the BH keep extracting the energy from the BH by the superradiance. 
This clearly indicates the growth of the cloud, which is called the superradiant instability.  

Owing to the superradiant instability, the condensate of axion, {\it i.e.}, an axion cloud, is spontaneously formed and grows. 
The growth rate of the cloud can be calculated by the matched asymptotic expansion \cite{Detweiler:1980uk} for the axion mass with $\mu M \ll 1$, and by the WKB method for $\mu M \gg 1$ \cite{Zouros:1979iw, Arvanitaki:2010sy}, or by the numerical calculation with the continued fraction method \cite{Dolan:2007mj, Yoshino:2015nsa}. 
The numerical results show the growth rate takes the maximum at $l=m=1, a/M\sim 1, \mu M \sim 0.42$ with $M \omega_I \sim 1.5 \times 10^{-7}$. 
The time scale for this growth is around 1 minute for a solar mass BH. 
This is much shorter than the age of the Universe, and hence axion clouds can become very heavy and dense.

\section{Adiabatic evolution of a self-interacting axion cloud}\label{sec:3}

In the preceding section, we saw that the superradiant instability is fast enough for the axion cloud to grow to be so dense that the self-interaction of the axion cannot be neglected. One of the most interesting possibility caused by the self-interaction is the bosenova, 
which is the collapse of clouds accompanied by a strong gravitational wave emission. 
Numerical simulation in \cite{Yoshino:2015nsa} and perturbative calculation in \cite{Omiya:2020vji} support the occurrence of the bosenova. 
However, it is still unclear whether the bosenova actually happens or not, 
since both methods cannot track the evolution of clouds starting with a small amplitude $(\phi \lesssim \mu/F_a)$ till the onset of the bosenova with a large amplitude $(\phi \sim 1)$. In the following, we directly solve the nonlinear equation of motion for the axion \eqref{eq:eomfull}
without truncating the potential to study the long-term evolution of an axion cloud. Below, we assume that the cloud starts with a state occupying only one single superradiant mode with $(l,m,\omega) = (l_0,m_0,\omega_0)$, for simplicity.

Our strategy is to use the fact that the cloud grows adiabatically, 
even when the cloud becomes so dense that the perturbative treatment of the self-interaction is not valid any more \cite{Omiya:2020vji}. 
Here, ``adiabatic" means that the growth rate of the cloud is much smaller than the dynamical time scale of the cloud, {\it i.e.},  
\begin{align}
	\omega_I \ll \omega_R~.
\end{align}
During the adiabatic evolution, the shape and the amplitude of the axion field only gradually change in time. 
Thus, the axion field configuration in a short time scale is almost stationary with an approximately fixed amplitude $A_0$. 
Since it is likely that the symmetry of configuration is preserved under the adiabatic evolution, we assume that the configuration of the axion field with an arbitrary amplitude $A_0$ can be  approximated by the one with an approximate helical symmetry as
\begin{align}\label{eq:adiabaticansatz}
	\phi(A_0) =& \sum_{n = 1}^{\infty}\sum_{l\ge n m_0}^{\infty} e^{-i n (\omega_0(A_0) t - m_0 \varphi)} \tilde{R}_{nl}(r;A_0) Y_{l nm_0}(\cos\theta) + {\rm c.c.}~.
\end{align}
Here, c.c. denotes the complex conjugate and $Y_{lm}(x)$ is defined as 
\begin{align}
	Y_{lm}(x) \equiv N^m_{l} P^m_l(x)~,
\end{align}
with
\begin{align}
 N^m_{l}=\sqrt{\frac{(l - m)! (2l+1)}{2(l+m)!}}~, 
\end{align}
where $P^m_{l}(x)$ is the associated Legendre polynomial. 
We understand that $Y_{lm}(x)$ is normalized to satisfy
\begin{align}
	\int_{-1}^{1} dx\, Y_{lm}(x) Y_{l'm}(x) = \delta_{ll'}~.
\end{align}
Here, we define $A_0$ as a parameter that specifies the amplitude of the fundamental 
mode at a large radius, {\it i.e.},
\begin{align}
	\tilde{R}_{1l_0}(r;A_0) &\to A_0 \frac{e^{- \sqrt{\mu^2 - \omega_0^2} r}}{r/M} \left(\frac{r}{M}\right)^{-M \frac{\mu^2 - 2 \omega_0^2}{\sqrt{ \mu^2 - \omega_0^2}}}(1 + \mathcal{O}(r^{-1}))~, & (r &\to \infty)~.
\end{align}
Notice that the fundamental frequency of the configuration, $\omega_0 = \omega_0(A_0)$, also depends on the amplitude $A_0$.

Substituting the ansatz \eqref{eq:adiabaticansatz} to the equation of motion \eqref{eq:eomfull} 
and neglecting the time derivative of the amplitude parameter $A_0$, we obtain
\begin{align}
	\OD{r}\left(\Delta\ODD{\tilde{R}_{nl}}{r}\right)  &+ \left[\frac{n^2(\omega_0 (r^2 + a^2) - a m_0)^2}{\Delta} - \mu^2 r^2 + 2 a n^2 \omega_0 m_0 - a^2 n^2 \omega_0^2  -l(l+1) \right. \cr
	&\qquad \left. + a^2 (n^2 \omega_0^2 - \mu^2)\frac{1-2l(l+1) + 2 n^2 m_0^2}{3-4l(l+1)} \right]\tilde{R}_{nl} \cr
	& + a^2 (n^2 \omega_0^2 - \mu^2)\left(\frac{(l-1-nm_0)(l-nm_0)}{(2l-3)(2l-1)}\frac{N^{nm_0}_{l-2}}{N^{nm_0}_{l}}\tilde{R}_{nl-2} \right.\cr
	&\qquad \left.+ \frac{(l+2 + nm_0)(l+1+nm_0)}{(2l+3)(2l+5)}\frac{N^{nm_0}_{l+2}}{N^{nm_0}_{l}}\tilde{R}_{nl+2} \right) \cr
	& + \int_{0}^{2\pi}\frac{d\varphi}{2\pi} \int _{-1}^1 dx \ Y_{lnm_0}(x) e^{-in m_0 \varphi} (r^2 + a^2 x^2)V'(\phi) = 0 \label{eq:Rlmeq}~.
\end{align}
We impose the ingoing boundary condition at the event horizon and the outgoing boundary condition at infinity. 
Since nonlinear terms fall off sufficiently fast for $r\to \infty$ and for $r \to r_+$, 
the asymptotic solutions satisfying these boundary conditions can be derived by neglecting the nonlinear terms, which are given by 
\begin{align}\label{eq:BClin}
	\tilde{R}_{nl} \to &  A^{(in)}_{nl}\left(\frac{r-r_-}{M}\right)^{i n\frac{2 M r_-}{r_+-r_-}(\omega_0 - m\frac{a}{2M r_-}) - (1-2nM\omega_0 i)- i M \frac{\mu^2 - 2 n^2 \omega_0^2}{\sqrt{n^2 \omega_0^2 - \mu^2}}} \cr
	&\times e^{i \sqrt{n^2\omega_0^2 - \mu^2} (r-r_-)} \left(\frac{r-r_+}{M}\right)^{- i n\frac{2 M r_+}{r_+ - r_-}(\omega_0 - m \Omega_H)}~, & (r &\to r_+)~,\\
	\label{eq:BClinH}
	\tilde{R}_{nl} \to & A^{(out)}_{nl}\frac{e^{+i \sqrt{n^2\omega_0^2 - \mu^2}r}}{r/M} \left(\frac{r}{M}\right)^{- i M\frac{\mu^2 - 2 n^2 \omega_0^2}{\sqrt{n^2 \omega_0^2 - \mu^2}}}\cr
	&\times\left(1 + \frac{a_1}{r/M} +\frac{a_2}{(r/M)^2} +\cdots \right) ~, & (r&\to \infty)~.
\end{align}
The coefficients $a_1, a_2, \cdots$ are analytically determined order by order, and we calculate up to $a_7$. 

Since Eq. \eqref{eq:adiabaticansatz} is valid only for a short period, much shorter than the  growth time scale $\ll \omega_{0I}^{-1}$, it cannot be a globally valid solution. 
To obtain a solution global in time, we demand the amplitude $A_0$ to slowly vary in time,  in such a way that the one-parameter family of solutions $\{\phi(A_0)\}_{A_0}$ is swept to satisfy the energy balance. Namely, the time dependence of $A_0$ is determined by
\begin{align}\label{eq:renomA0}
	\ODD{E(A_0)}{A_0} \ODD{A_0}{t} = -F_{\rm tot}(A_0)~.
\end{align}
Here, $E(A_0)$ and $F_{\rm tot}(\phi(A_0))$ are, respectively, the energy and the net energy flux of the quasi-stationary configuration \eqref{eq:adiabaticansatz} with a given amplitude $A_0$. Using energy-momentum tensor $T_{\mu\nu}(A_0)$ of the axion configuration \eqref{eq:adiabaticansatz}, we obtain
\begin{gather}
	E(A_0) = \int dr\, d\cos\theta\, d\varphi \ (r^2 + a^2 \cos^2\theta)\sqrt{g^{tt}}T_{\mu\nu}(A_0)\xi^{\mu}_{(t)} n^\nu_{(t)}~,\\
	F_{\rm tot}(A_0)  = F_{H}(A_0) + F_{{\infty}}(A_0)~,\\
	F_H(A_0) =  \int d\cos\theta d\varphi\, 2 M r_+ T_{\mu\nu}(A_0)\xi^{\mu}_{(t)} l^{\nu}|_{r=r_+}~,\\
	F_{\infty}(A_0) = \int d\cos\theta d\varphi\, (r^2 + a^2 \cos^2\theta)\sqrt{g^{rr}}T_{\mu\nu}(A_0)\xi^{\mu}_{(t)} n^{\nu}_{(r)}|_{r \to \infty}~.
\end{gather}
Here, $\xi^{\mu}_{(t)} = (1,0,0,0)$ and $l^{\mu} = \frac12(1,\Delta/(r^2 + a^2),0,a/(r^2 + a^2))$ in the BL coordinates \cite{1974ApJ...193..443T}. In addition, $n^\mu_{(t)}$ and $n^\mu_{(r)}$ are unit vectors normal to the $t = {\rm constant}$ and $r = {\rm constant}$ surfaces, respectively. The explicit expressions are given by
\begin{align}
	\sqrt{g^{tt}}T_{\mu\nu}\xi^{\nu}_{(t)} n^\nu_{(t)} = {T^t}_{t}  =& \frac{(r^2 + a^2)^2 - \Delta a^2 \sin^2\theta}{2\rho^2 \Delta}(\partial_t \phi(A_0))^2 + \frac{\Delta - a^2 \sin^2\theta}{2 \rho^2 \Delta \sin^2\theta}(\partial_\varphi \phi(A_0))^2 \cr
	&+ \frac{\Delta}{2\rho^2}(\partial_r \phi(A_0))^2 + \frac{1}{2\rho^2}(\partial_\theta \phi(A_0))^2 + V(\phi(A_0))~,\\
	T_{\mu\nu}\xi^{\mu}_{(t)} l^{\nu}|_{r=r_+} =& (\partial_t\phi(A_0) + \Omega_H \partial_\varphi\phi(A_0))\partial_t\phi(A_0)~,\\
	(r^2 + a^2 \cos^2\theta)\sqrt{g^{tt}}& T_{\mu\nu}\xi^{\mu}_{(t)} n^{\nu}_{(r)}|_{r \to \infty} = (r-r_+)(r-r_-)\partial_r\phi(A_0) \partial_t\phi(A_0)~.
\end{align}
The angular momentum of the cloud $J_{cl}(A_0)$ can be calculated similarly to the energy of the cloud as 
\begin{align}\label{eq:angularmom}
    J_{cl}(A_0) = \int dr\, d\cos\theta\, d\varphi \ (r^2 + a^2 \cos^2\theta)\sqrt{g^{tt}}T_{\mu\nu}(A_0)\xi^{\mu}_{(\varphi)} n^\nu_{(t)}~,
\end{align}
where $\xi^{\mu}_{(\varphi)}= (0,0,0,1)$. When the non-linearlity is small, this expression can be approximated as 
\begin{align}
    J_{cl}(A_0) \sim \frac{m_0}{\omega_{0R}(A_0)} E_{cl}(A_0)~.
\end{align}

\section{Numerical Result}\label{sec:4}

Here, we present numerical results obtained by the method explained in the previous section. 
We obtain a sequence of solutions of Eq. \eqref{eq:Rlmeq} by gradually increasing the amplitude $A_0$ starting with a small value, where the linear approximation is a good approximation. 
Below, we focus on the fastest growing mode with $l_0 = m_0 = 1$. Numerical calculation is performed with {\it Mathematica}. Details of our numerical calculation method are explained in Appendix \ref{app:A}.

\subsection{The evolution of a cloud for the fastest growing parameter set}
\label{sec:4.1}

\begin{figure}[t]
\begin{tabular}{cc}
\begin{minipage}[t]{0.5\hsize}
	\centering
	\includegraphics[keepaspectratio,scale=0.21]{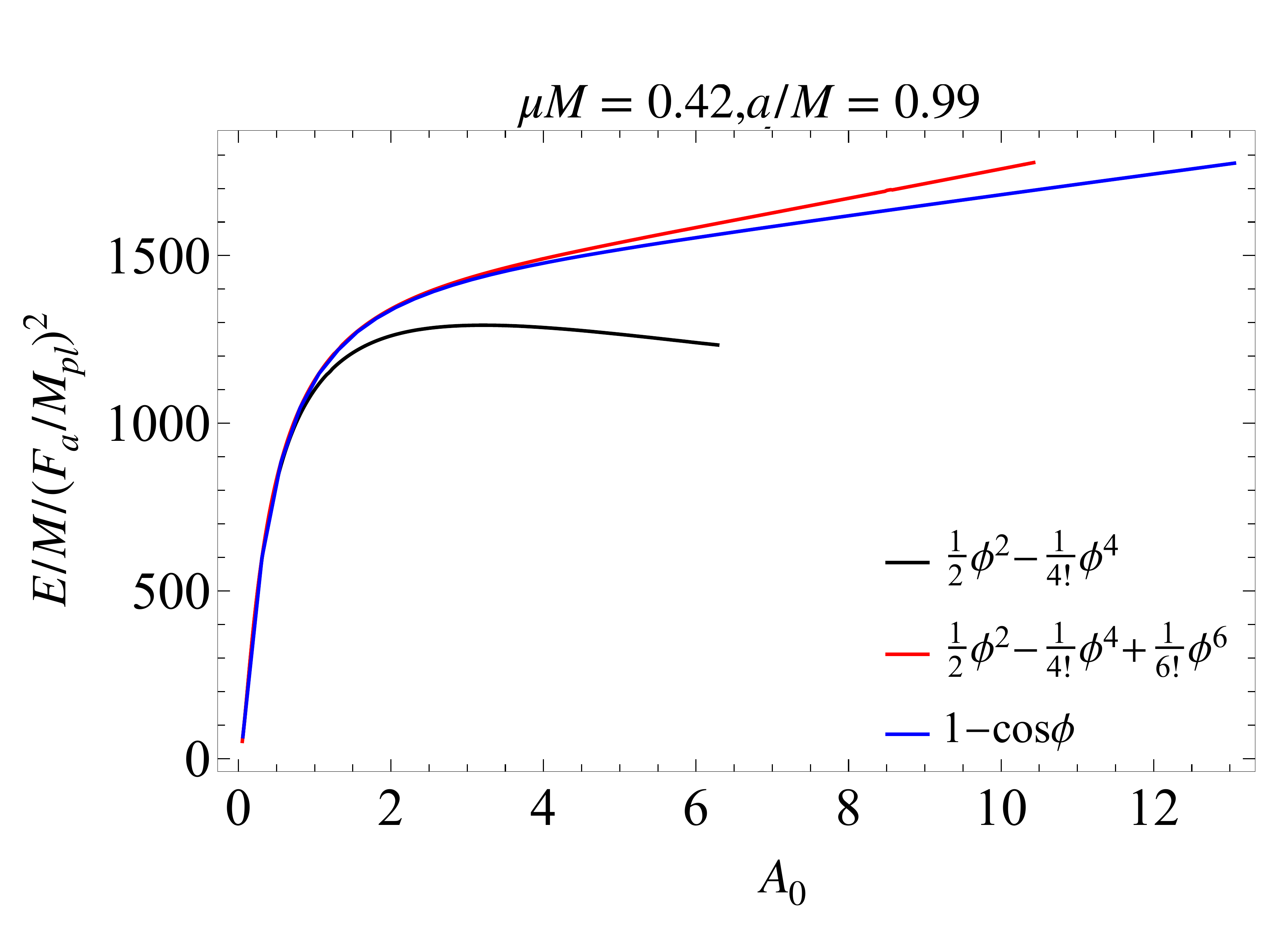}
\end{minipage} &
\begin{minipage}[t]{0.5\hsize}
	\centering
	\includegraphics[keepaspectratio,scale=0.21]{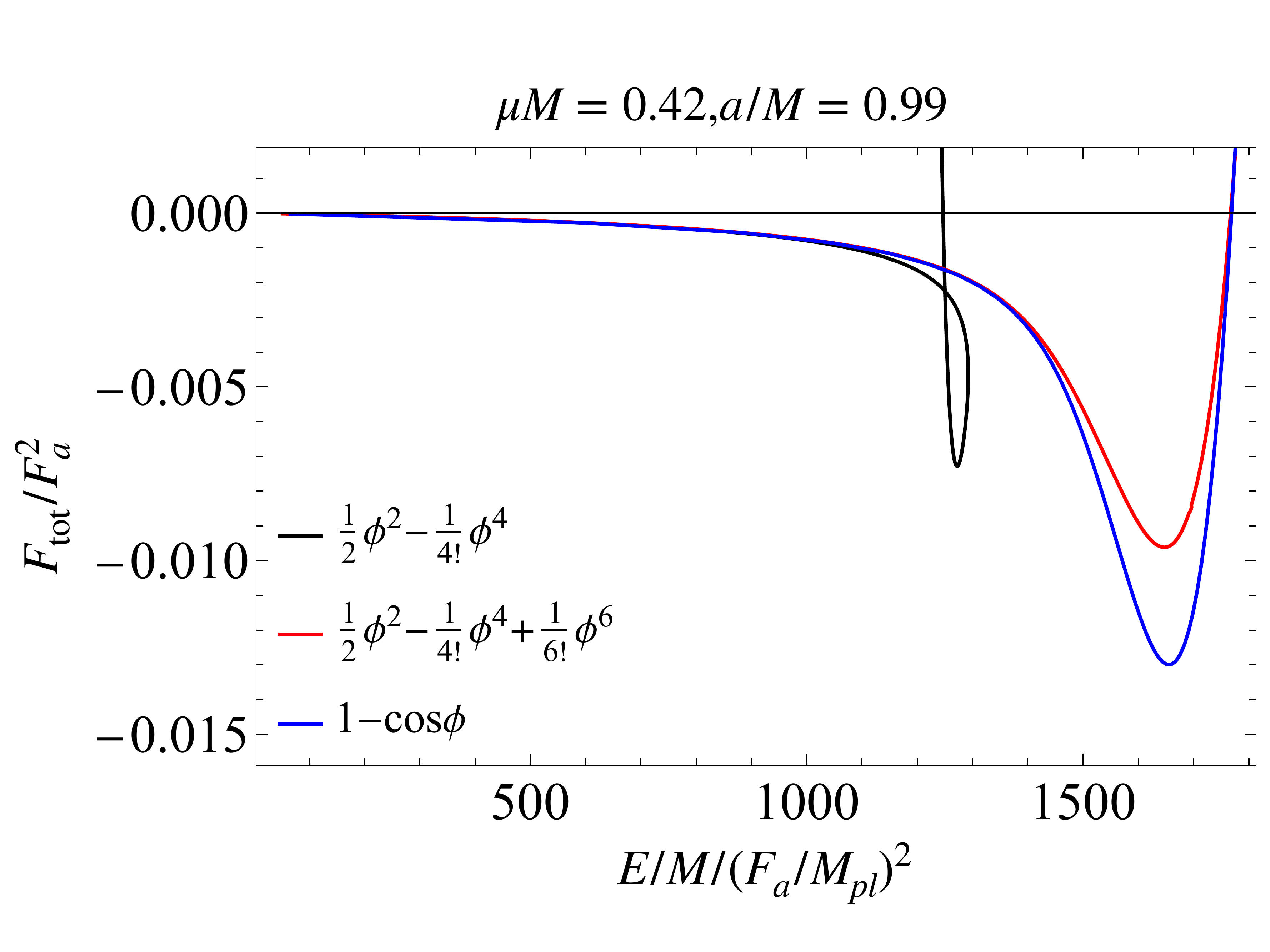}
\end{minipage}
\end{tabular}
	\caption{(Left) Dependence of the energy $E(A_0)$ on the amplitude $A_0$. The black, red, and blue curves correspond to the cases with the potential given in Eqs. \eqref{eq:phi4}, \eqref{eq:phi6}, and \eqref{eq:cos}, respectively.  (Right) Dependence of the total flux $F_{\rm tot}(A_0)$ on the energy $E$. In the same way as the left panel, the black, red, and blue curves correspond to the respective choices of the potential. }
	\label{fig:energyandflux}
\end{figure}

First, we show the result with $a/M = 0.99$ and $\mu M = 0.42$, which gives a growth rate quite close to the maximum value \cite{Dolan:2007mj}. In this subsection, we consider the following three different potentials,  
\begin{align}
    \label{eq:phi4}
	V(\phi) &= \mu^2\left(\frac{1}{2}\phi^2 - \frac{1}{4!}\phi^4\right)~,\\
	\label{eq:phi6}
	V(\phi) &= \mu^2\left(\frac{1}{2}\phi^2 - \frac{1}{4!}\phi^4 + \frac{1}{6!}\phi^6\right)~,\\
	\label{eq:cos}
	V(\phi) &= \mu^2 \left(1 - \cos\phi\right)~,
\end{align}
to see the influence of the higher order terms of the potential on the evolution. 

In Figs. \ref{fig:energyandflux}, we show how the energy $E$ and the total flux $F_{\rm tot}$ depend on the amplitude $A_0$. 
As one can see from the left panel of Fig. \ref{fig:energyandflux}, the energy with an attractive $\phi^4$ interaction \eqref{eq:phi4} takes a maximum, $dE/dA_0 = 0$, at $A_0 \sim 3$. 
The appearance of a maximum corresponds to the presence of a neutral perturbation at this amplitude, which would indicate the onset of dynamical instability \cite{1983bhwd.book.....S}. 
Namely, the axion cloud with the attractive $\phi^4$ potential will become unstable at this amplitude. 
By contrast, the clouds with the potential \eqref{eq:phi6} or \eqref{eq:cos} never become unstable. 
This is because the repulsive interaction from the higher order terms in the cosine type potential stabilizes the cloud. 

\begin{figure}[t]
	\centering
	\includegraphics[keepaspectratio,scale=0.3]{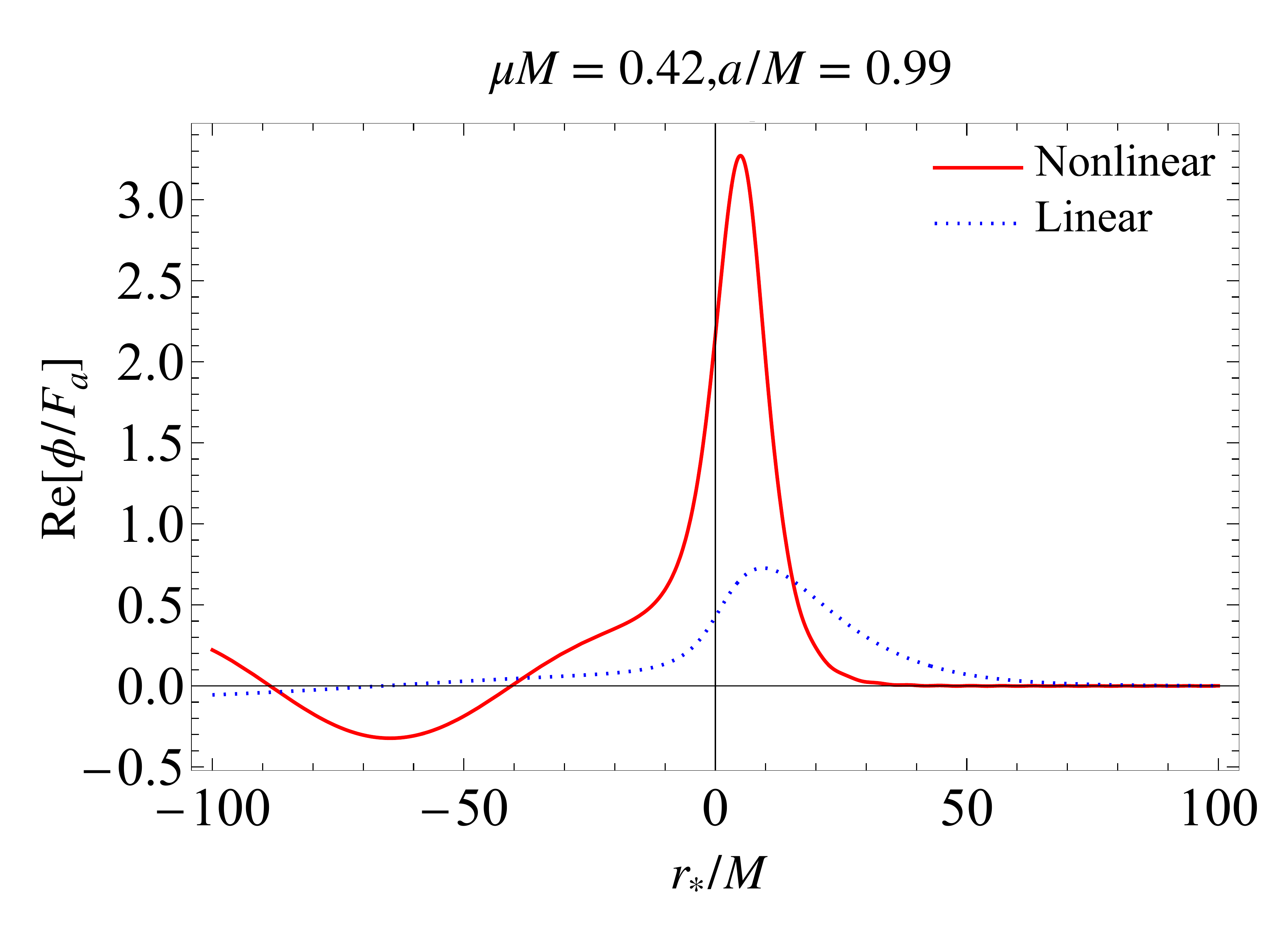}
	\caption{The red and blue dotted curves, respectively, show the configuration of the nonlinear quasi-stationary cloud with the cosine potential \eqref{eq:cos} and that of the linear cloud (Eq. \eqref{eq:sep}) on the equatorial plane. In both cases the energy is fixed to $E/M = 1768 (F_a/M_{pl})^2$, where the growth of the amplitude due to the superradiance saturates. Here, we explicitly write $M_{pl}^2=G^{-1}=1$ to make it clear that $F_a$ should be counted in the Planck unit.}
	\label{fig:fieldconfig042}
\end{figure}

As the amplitude of the cloud becomes large, the energy flux to infinity increases, and eventually it balances with the energy gain due to the superradiance (right panel of Fig. \ref{fig:energyandflux}). 
Since the cloud with the potential \eqref{eq:phi6} and \eqref{eq:cos} is stable throughout the evolution, as the final state a quasi-stationary state would be realized. This saturation occurs at $A_0 \sim 12$ or $E/M \sim 1.8\times 10^{3} (F_a/M_{pl})^2$, which is only $0.1 \%$ of the BH mass if we choose the decay constant to be the GUT scale, $F_a \sim 10^{16} {\rm GeV}$. 
In Fig. \ref{fig:fieldconfig042}, we show the configuration of the axion cloud in this quasi-stationary state and the linear configuration given by Eq. \eqref{eq:sep} with $n=0$. We normalize the linear configuration to have the same energy as the nonlinear quasi-stationary state.
We observe that the nonlinear quasi-stationary state is more compact than the configuration of the linear solution.  
This is a consequence of the attractive force originating from the leading $\phi^4$ interaction term in the cosine-type potential.

\begin{figure}[t]
\begin{tabular}{cc}
\begin{minipage}[t]{0.5\hsize}
	\centering
	\includegraphics[keepaspectratio,scale=0.22]{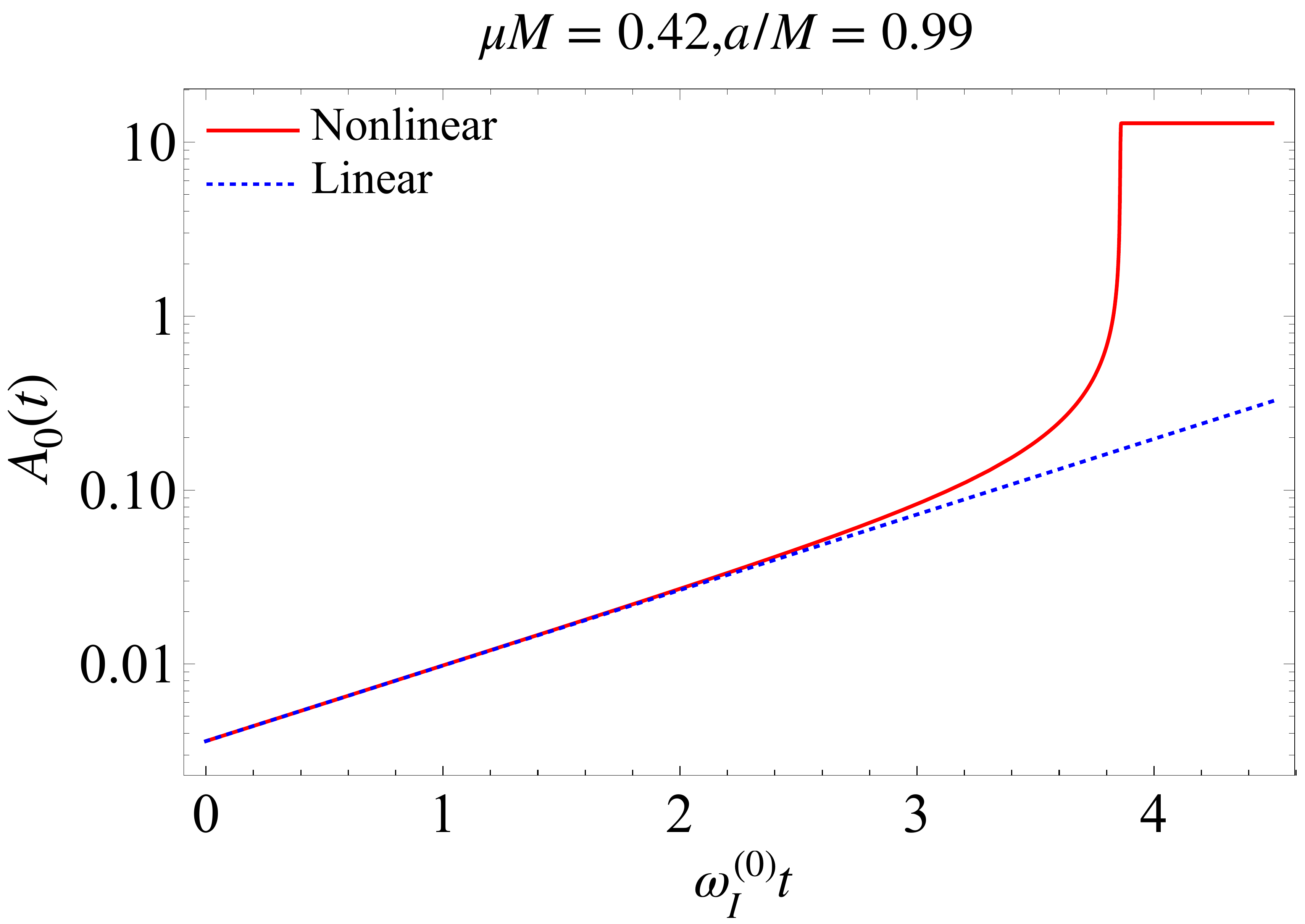}
\end{minipage} &
\begin{minipage}[t]{0.5\hsize}
	\centering
	\includegraphics[keepaspectratio,scale=0.22]{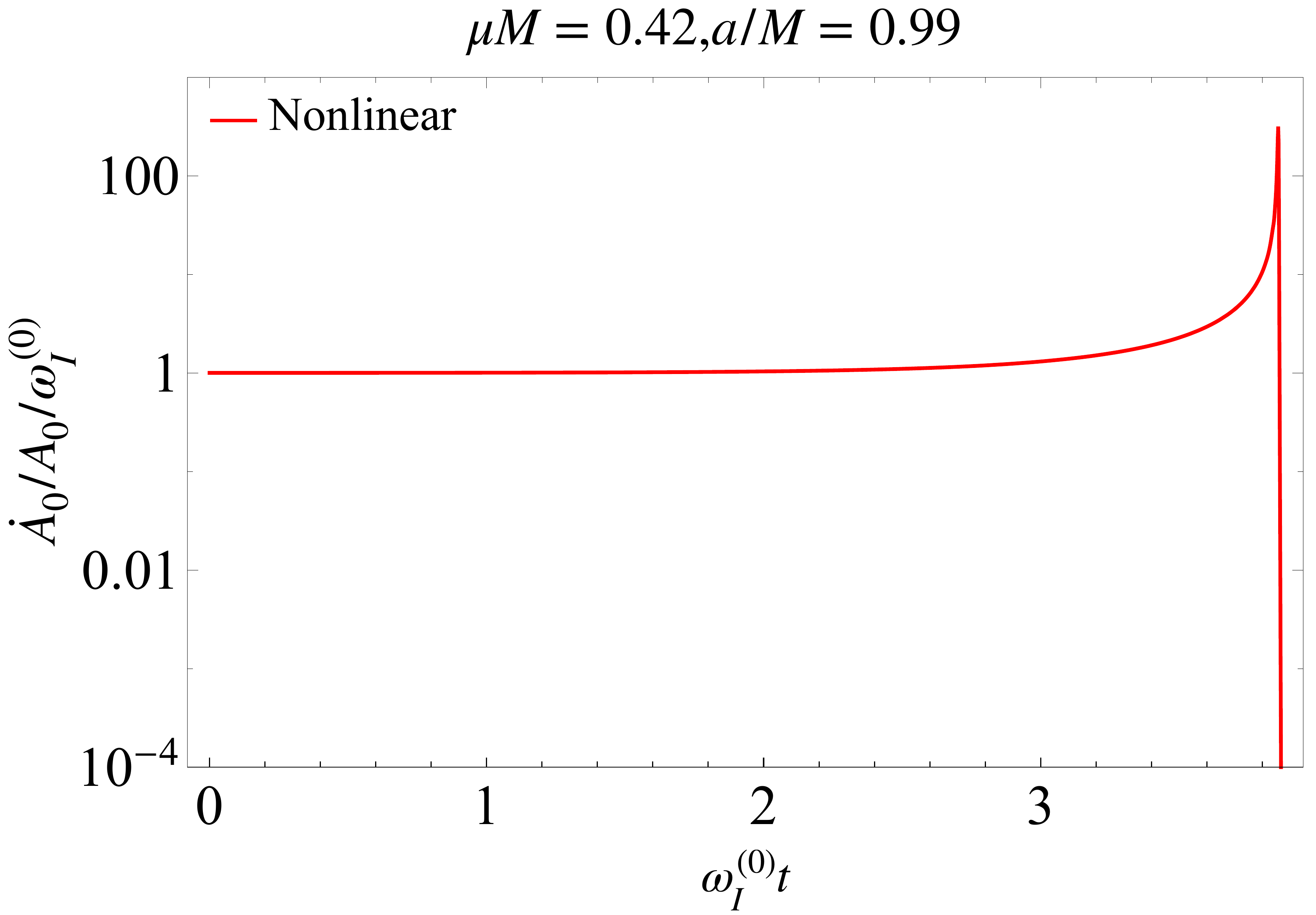}
\end{minipage}
\end{tabular}
	\caption{(Left) Red curve and the blue dotted straight line show the nonlinear (cosine-type potential \eqref{eq:cos}) and the linear time evolutions of the amplitude $A_0$,  respectively. The horizontal axis is the normalized time $\omega^{(0)}_I t$, where $\omega^{(0)}_I$ is the growth rate in the linearized model, not the imaginary part of the frequency $\omega_0$. (Right) The inverse time scale of the cloud evolution $\dot{A}_0/A_0$, normalized by $\omega^{(0)}_I$. Again, the horizontal axis is the normalized time, $\omega^{(0)}_I t$.}
	\label{fig:ampevol}
\end{figure}

\begin{figure}[t]
	\centering
	\includegraphics[keepaspectratio,scale=0.35]{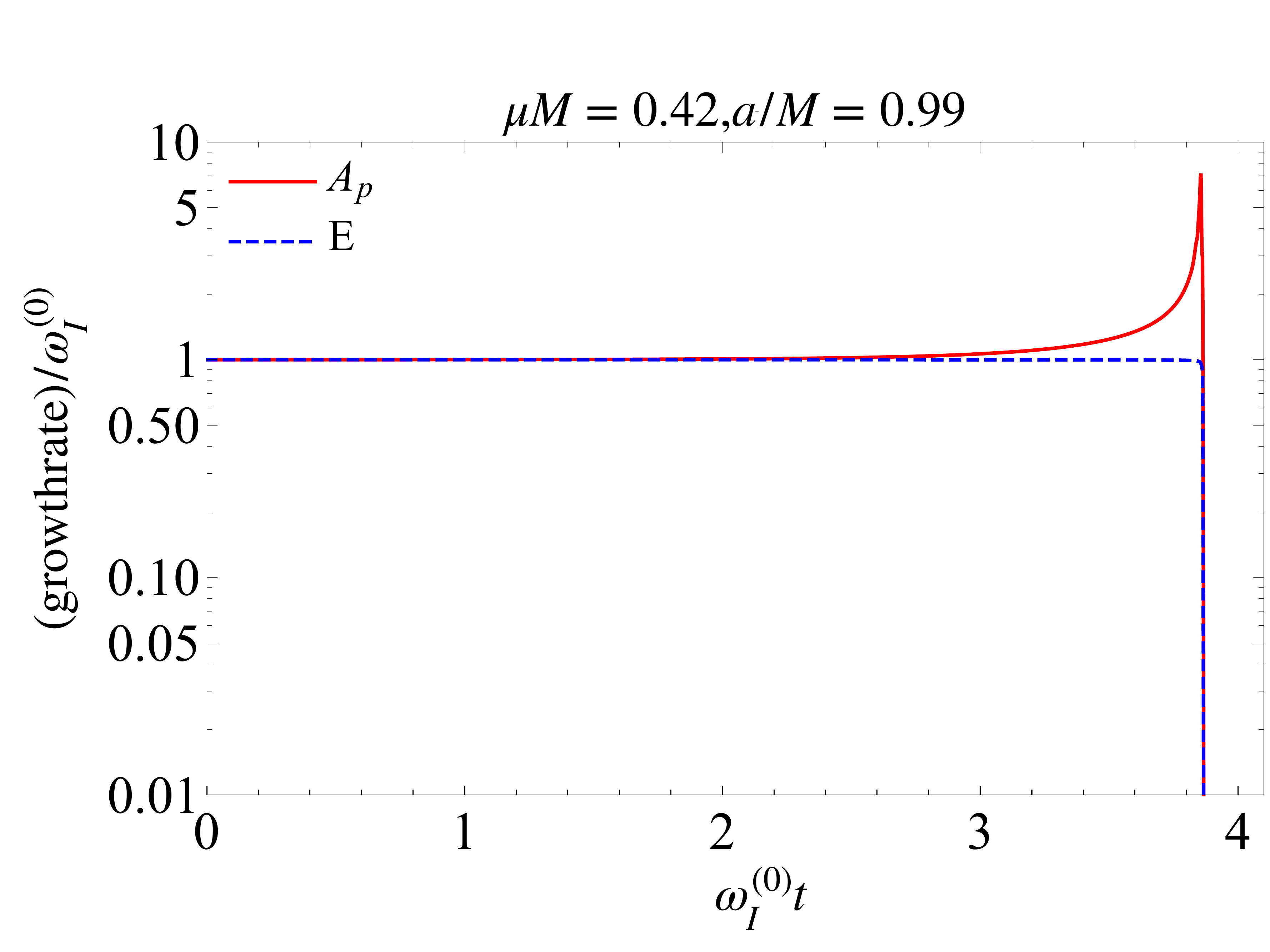}
	\caption{Red solid (blue dotted) curve corresponds to the time evolution of the growth rate defined by $\dot{A}_p/A_p$ ($\dot{E}/2E$) divided by the growth rate determined by the imaginary part of the frequency,  $\omega_{0I}$. Here we emphasize that $\omega_{0}$ is the fundamental frequency of the nonlinear configuration which is different from the frequency in the linearlized model, $\omega^{(0)}$.}
	\label{fig:growthcomp}
\end{figure}
Now, we know the dependences of the energy and the energy flux on the amplitude, we can calculate the time evolution of the amplitude via Eq. \eqref{eq:renomA0}. In the left panel of Fig. \ref{fig:ampevol}, we show the time evolution of the amplitude. 
We observe that the growth is accelerated in the early time. 
After the amplitude becomes large enough, the energy emission to infinity balances with the superradiant growth, and the evolution of the amplitude terminates. 
The acceleration is due to the attractive nature of the leading interaction term $\propto \phi^4$ in the cosine-type potential. 
In the right panel of Fig. \ref{fig:ampevol}, we show the time scale of the cloud evolution. The growth rate of a nonlinear cloud becomes about 100 times larger than the original growth rate. Nevertheless, the original growth rate is much smaller than the dynamical time scale by $10^{-7}$.  
Hence, the growth rate even after the cloud becomes nonlinear is still much smaller than the inverse of the dynamical time scale. This confirms the validity of the adiabatic evolution assumed in our scheme and thus that of the sequence of quasi-stationary states obtained by our calculation.

To further check the consistency of our calculation scheme, we show the growth rate of the peak amplitude and 
that of the energy. 
In Fig. \ref{fig:growthcomp}, we show the time evolution of these growth rates divided by the imaginary part of the frequency $\omega_{0I}$. 
The deviation of these ratios from unity remains to be at most $\mathcal{O}(1)$. In other words, difference between the growth rates defined in different ways are $\mathcal{O}(\omega_{0I})$, which is tiny. 
This difference represents the error due to our naive ansatz on the time dependence of each mode. 
As we have confirmed that the error when we substitute the solution with the time-dependent magnitude $A_0(t)$ into the equation of motion is suppressed by $\omega_{0I}$, 
we can conclude that the higher order correction to amend this error is tiny. 
It would be interesting to point out that the growth rate determined by the time evolution of the 
energy, $\dot E/2E$, remains to be very close to $\omega_{0I}$, even when the configuration becomes nonlinear.

\subsection{Dependence on the axion mass and black hole spin}\label{sec:4.2}

\begin{table}[t]
  \caption{Parameters $a/M$ and $\mu M$ we have calculated. The corresponding frequency $\omega_{0R}$ and the growth rate $\omega_{0I}$ of the superradiant mode derived by the linearized equation of motion are also shown.}
  \label{table:calculatedmodes}
  \centering
  \begin{tabular}{lccccc}
    \hline
    $(a/M,\mu M)$  & $M \omega_{0R}$  &  $M \omega_{0I}$\\
    \hline \hline
    (0.99,0.42)  & 0.4088  & $1.504\times 10^{-7}$ \\
    (0.99,0.29)  & 0.2867  & $2.154\times 10^{-8}$ \\
    (0.9,0.29)  & 0.2867  & $1.543\times 10^{-8}$ \\
    (0.99,0.15)  & 0.1496  & $1.837\times 10^{-10}$ \\
    (0.9,0.15)  & 0.1496  & $1.737\times 10^{-10}$ \\
    (0.7,0.15)  & 0.1496  & $1.154\times 10^{-10}$ \\
    \hline
  \end{tabular}
\end{table}

Next, we change the axion mass $\mu M$ and the BH spin $a/M$, to see the effect of the variation of these parameters on the evolution. The parameter sets that we present in this paper are shown in Table \ref{table:calculatedmodes}. In the following, we fix the potential of the axion to the cosine-type one \eqref{eq:cos}.

\begin{figure}[t]
\begin{tabular}{cc}
\begin{minipage}[t]{0.5\hsize}
	\centering
	\includegraphics[keepaspectratio,scale=0.21]{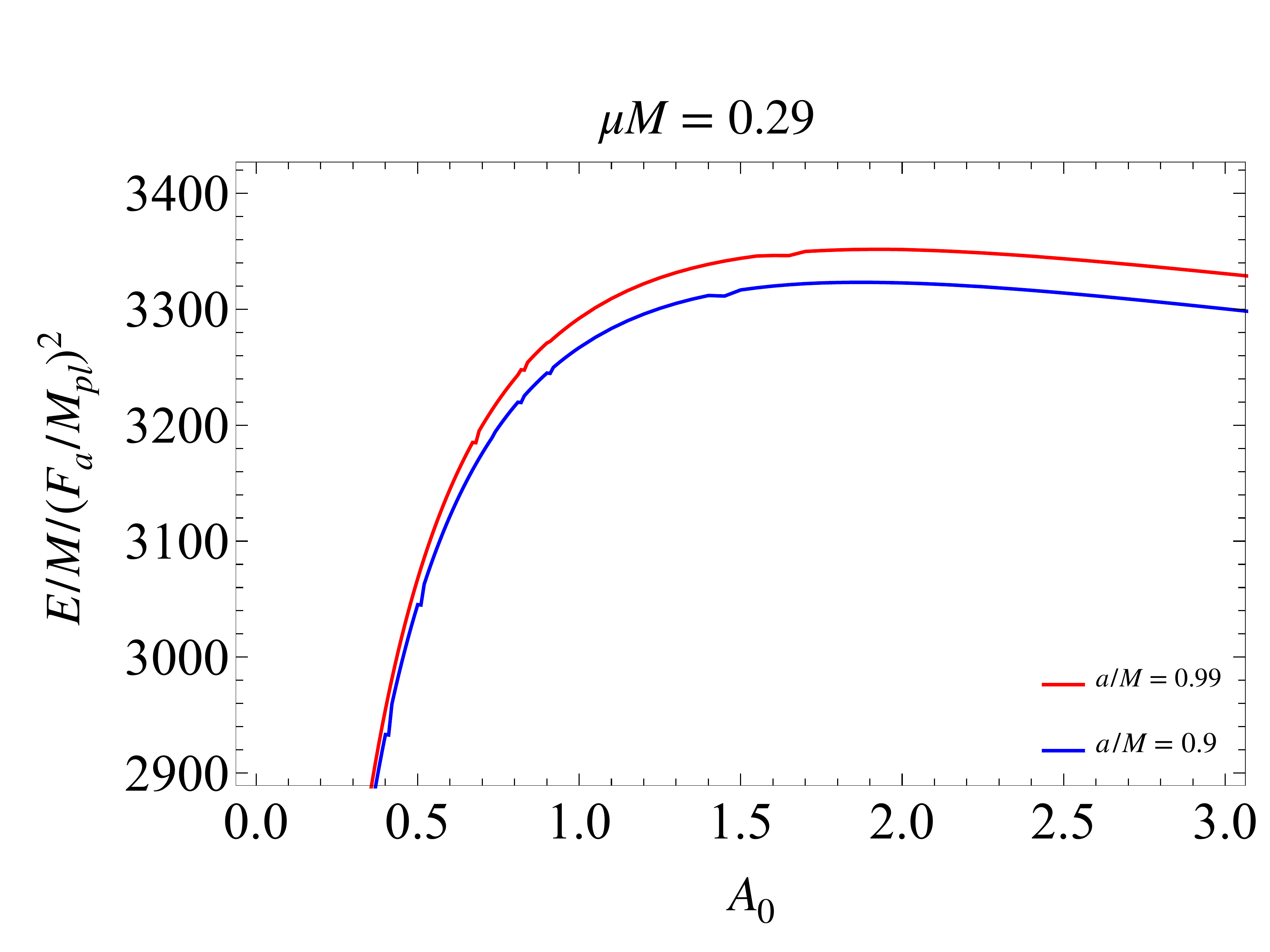}
\end{minipage} &
\begin{minipage}[t]{0.5\hsize}
	\centering
	\includegraphics[keepaspectratio,scale=0.21]{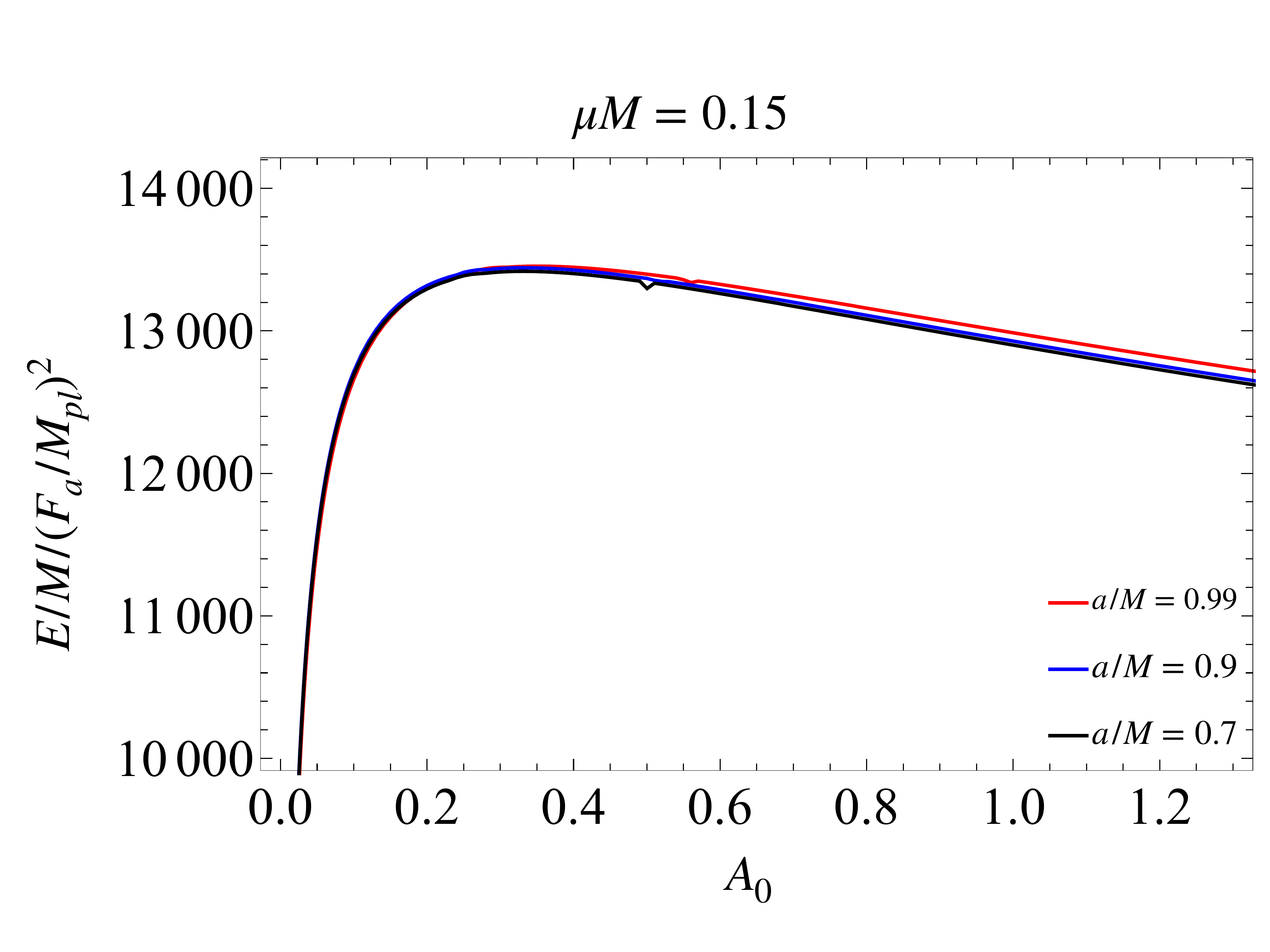}
\end{minipage}
\end{tabular}
	\caption{(Left) Dependence of the energy divided by $(F_a/M_{pl})^2$ on the amplitude $A_0$ for $\mu M = 0.29$ case. The red and blue curves correspond to the $a/M=0.99$ and $a/M=0.9$ cases, respectively. (Right) The same figure but with $\mu M = 0.15$. The red, blue, and black curves correspond to the $a/M=0.99, a/M=0.9,$ and $a/M=0.7$ cases, respectively.}
	\label{fig:nonrelenergy}
\end{figure}

We start with looking at the dependence of the energy on the amplitude $A_0$, as in the previous subsection. Figure \ref{fig:nonrelenergy} shows the dependence of the energy on the amplitude $A_0$ for $\mu M = 0.29$ (left) and $\mu M = 0.15$ (right) cases. We observe that there exists a local maximum regardless of the value of $a/M$. Thus, if $\mu M$ is not so large, we find that the cloud always becomes unstable for any spin. 
The difference between the case with $\mu M =0.42$ and the case with $\mu M = 0.29$ or $0.15$ would be qualitatively explained by the toy model presented in the next section. 

\begin{figure}[t]
\begin{tabular}{cc}
\begin{minipage}[t]{0.5\hsize}
	\centering
	\includegraphics[keepaspectratio,scale=0.21]{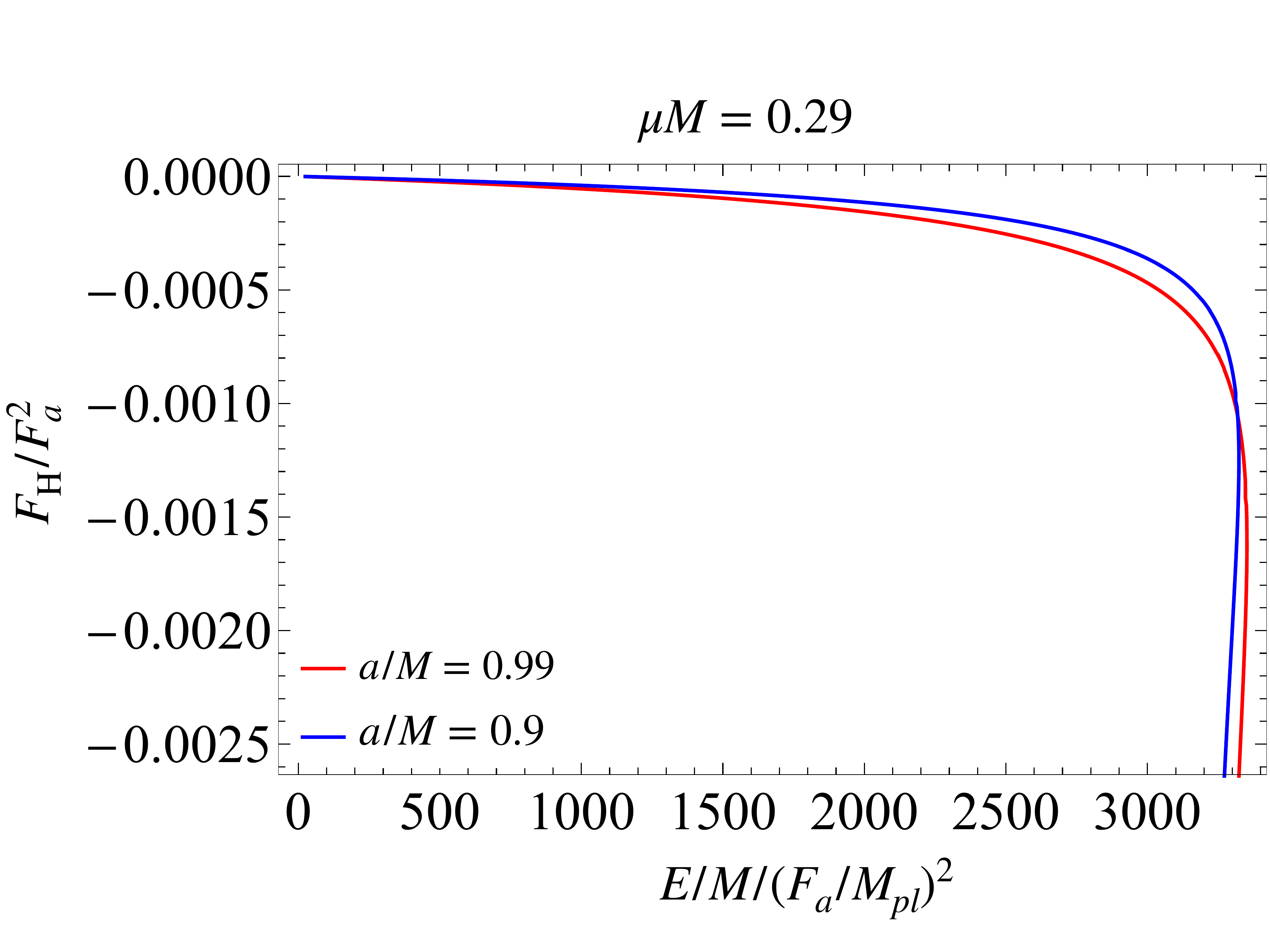}
	\includegraphics[keepaspectratio,scale=0.21]{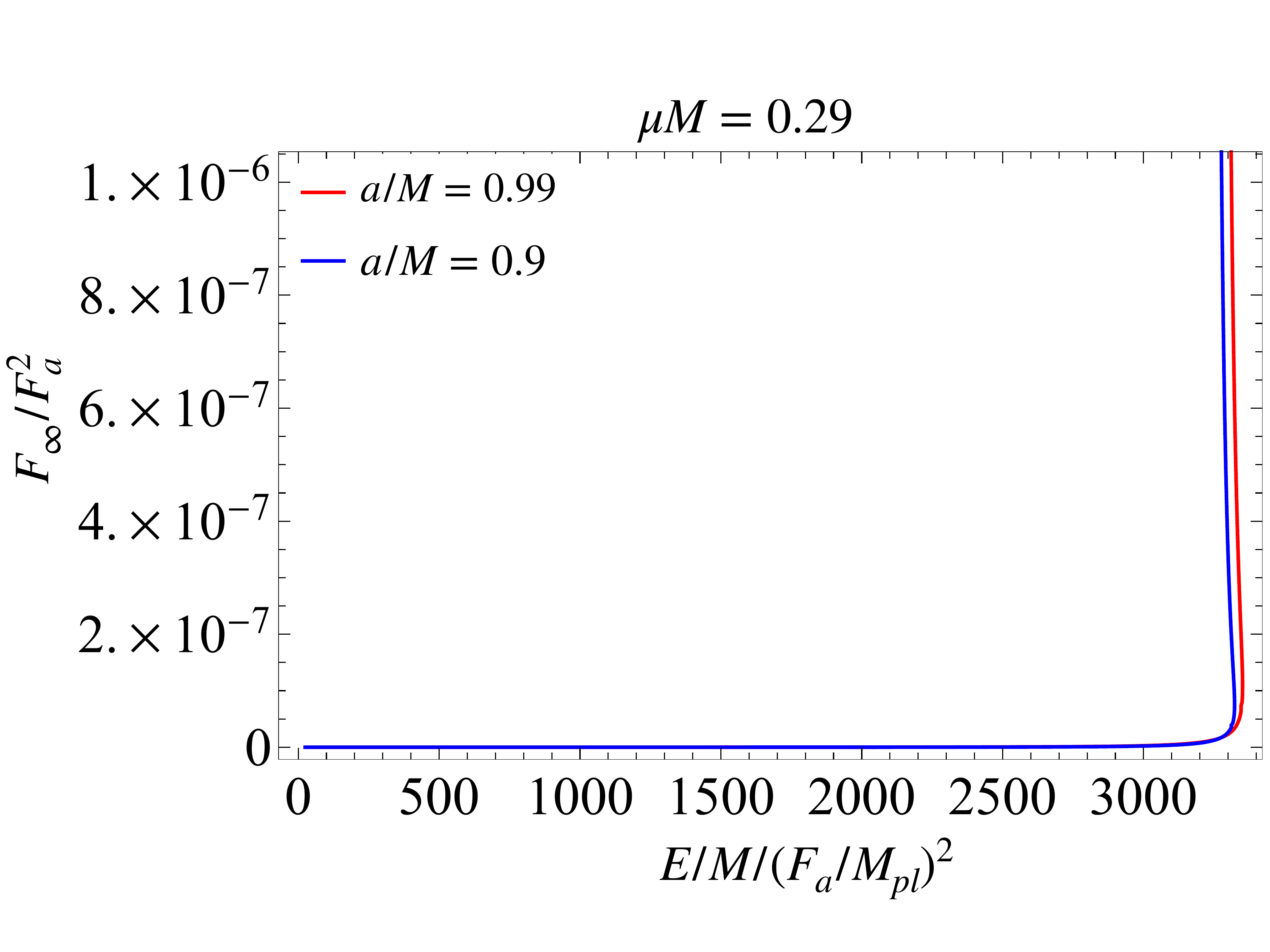}
\end{minipage} &
\begin{minipage}[t]{0.5\hsize}
	\centering
	\includegraphics[keepaspectratio,scale=0.21]{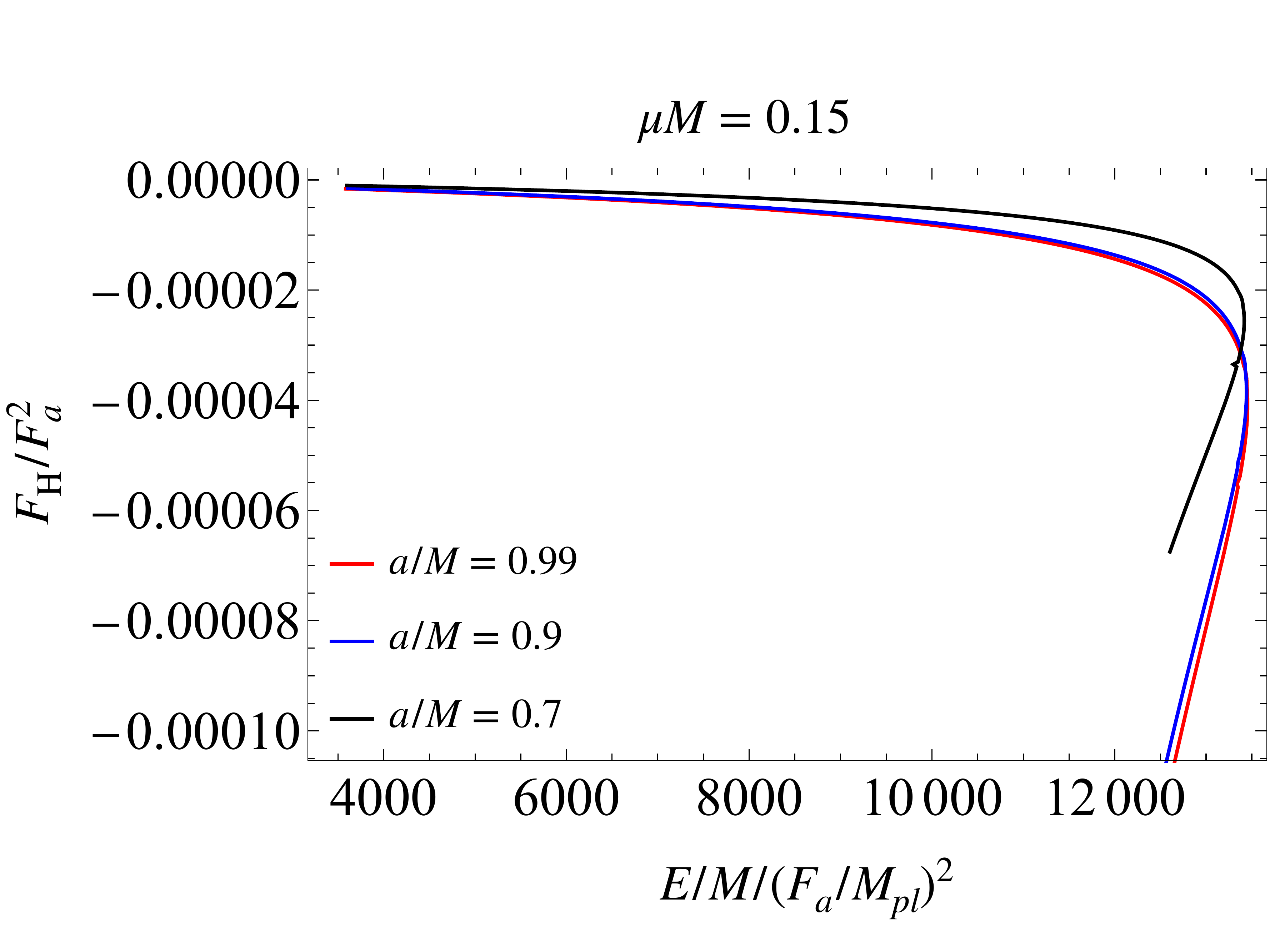}
	\includegraphics[keepaspectratio,scale=0.21]{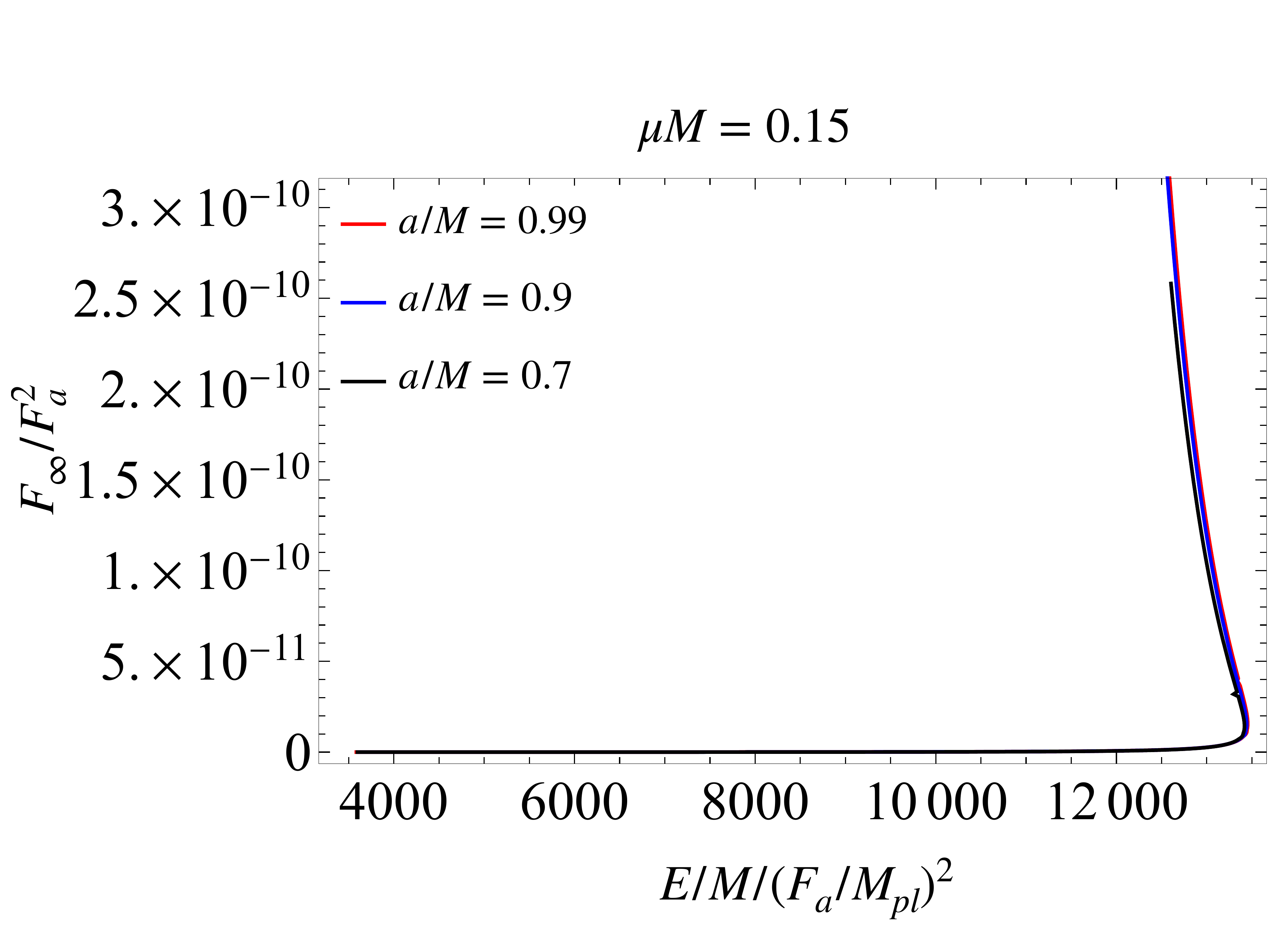}
\end{minipage}
\end{tabular}
	\caption{(Left) Dependence of the energy flux at horizon (upper panel) and infinity (lower panel) on the energy $E$ for $\mu M = 0.29$. Red and blue curves correspond to $a/M = 0.99$ and $0.9$ cases, respectively. The energy flux is normalized by $(F_a/M_{pl})^2$. (Right) The same graph but with $\mu M =0.15$. 
	The red, blue, and black curves correspond to the $a/M=0.99, a/M=0.9,$ and $a/M=0.7$ cases, respectively.}
	\label{fig:nonrelenergyflux}
\end{figure}

If the growth of the cloud saturates before the cloud becomes unstable, 
the instability will not happen. 
In Fig. \ref{fig:nonrelenergyflux}, we compare the energy flux from the horizon and that to infinity for each set of parameters ($\mu M =0.29$ on the left and $\mu M =0.15$ on the right), to see whether the dissipation to infinity terminates the growth or not. 
We find that the flux to infinity is largely suppressed, compared to the flux from the horizon, even when the energy of the cloud is large enough for the instability to occur. 
This is because in the small $\mu M$ limit, the wavelength of the outgoing wave of the axion with $m=3$ is around $\sim 1/3\mu$, which is much smaller than the size of the cloud $\sim M/(M\mu)^2$. 
Therefore, the dissipation to infinity is not efficient and cannot terminate the growth of the cloud 
before the onset of instability. 

\section{A toy model of the axion cloud}\label{sec:5}

In this section, we give a toy model which describes the qualitative behavior of the axion cloud seen in our numerical calculations. 
Our model is a simplified version of the effective theory proposed in \cite{Yoshino:2012kn}. 
In this model, we adopt the non-relativistic approximation, which keeps the leading terms in the expansion with respect to $\mu M$ and neglects the spin of the BH, assuming the form of the axion field as
\begin{align}
    \phi = \frac{1}{\sqrt{2\mu}}\left(\psi e^{- i \mu t} + \psi^* e^{+ i\mu t}\right)~,
\end{align}
and demanding that the characteristic length scale of the function $\psi$ is much longer than the Compton wavelength of the axion $\mu^{-1}$.

We take our starting point to be the action \eqref{eq:action} with the potential \eqref{eq:cospotential}. Under the non-relativistic approximation, the action takes the form of
\begin{align}
    S_{\rm NR} = F_a^2 \int dt\, d^3\!\bm{x}\ \left(\frac{i}{2}\left(\psi^* \dot{\psi} - \psi \dot{\psi}^*\right) - \frac{1}{2\mu}\left\vert \partial_i\psi\right\vert^2 + \frac{\mu M}{r}|\psi|^2 + \mu^2 \sum_{n=2}\frac{(-1/2)^n}{(n!)^2}\frac{|\psi|^{2n}}{\mu^n}\right)~.
\end{align}
From this non-relativistic action, we read the potential energy of the cloud as 
\begin{align}\label{eq:nonrelpot}
    V = \int d^3\!\bm{x}\ \left(\frac{1}{2\mu}|\partial_i\psi|^2 - \frac{\mu M}{r}|\psi|^2 - \mu^2 \sum_{n=2}\frac{(-1/2)^n}{(n!)^2}\frac{|\psi|^{2n}}{\mu^n}\right)~.
\end{align}
From our numerical calculation, we know that the configuration of the cloud is well approximated by a single spherical harmonics. Thus, we take an ansatz
\begin{align}\label{eq:nonrelansatz}
    \psi = A_p e^{-\frac{(r-r_p)^2}{4\sigma^2}} Y_{l_0m_0}(x)e^{+im_0\varphi}~,
\end{align}
for the cloud configuration. 
This wave function is characterized by the peak amplitude $A_p$, 
the position of the peak radius $r_p$, and the radial extension of the cloud $\sigma$. We plug in the ansatz \eqref{eq:nonrelansatz} into Eq. \eqref{eq:nonrelpot} and setting $l_0=m_0=1$, we obtain
\begin{align}\label{eq:potfixN}
    \frac{V}{N} =& \frac{r_p^2 + 3 \sigma^2}{8 \mu \sigma^2 (r_p^2 + \sigma^2)} + \frac{1}{\mu (r_p^2 + \sigma^2)} - \frac{\mu M r_p}{r_p^2 + \sigma^2}\cr
    &- \mu^2 \left(\frac{N_*}{160\pi \sqrt{2\pi}\mu^4 \sigma (r_p^2 + \sigma^2)} - \frac{3 N_*^2}{17920\pi^3 \mu^7 \sigma^2 (r_p^2 + \sigma^2)^2} + \cdots\right)~.
\end{align}
Here, $N$ is the particle number in the cloud defined by
\begin{align}\label{eq:nonrelpartnum}
    N= \int d^3\!\bm{x} \ |\psi|^2 \sim 2\pi \sqrt{2\pi}\sigma (r_p^2 + \sigma^2)A_p^2~,
\end{align}
where we ignore the inner cutoff of the radial integration and we define the dimensionless quantity $N_* \equiv \mu^2 N$\ \footnote{Since we have scaled the axion field $\phi$ by  the decay constant $F_a$, the correct particle number is given by $F_a^2 N$. Here, we defined $N_*$ by multiplying $N$ by $\mu^2$ instead of $F_a^2$ to eliminate $F_a$ from the potential, for simplicity.}. 
The radial integrations in Eq. \eqref{eq:nonrelpot} are also approximated as is performed in Eq. \eqref{eq:nonrelpartnum}. 

\begin{figure}[t]
\begin{tabular}{cc}
\begin{minipage}[t]{0.5\hsize}
	\centering
	\includegraphics[keepaspectratio,scale=0.21]{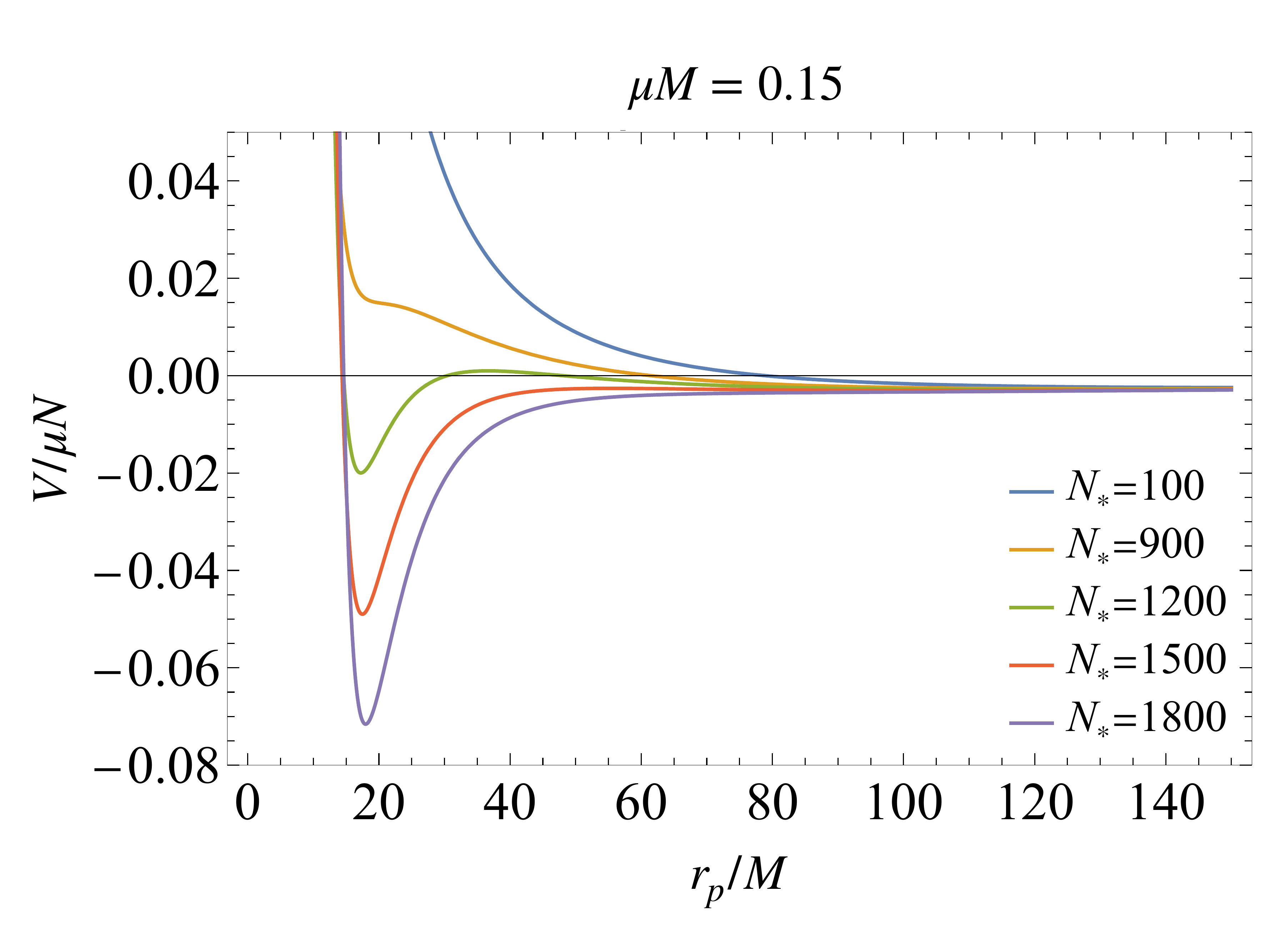}
\end{minipage}&
\begin{minipage}[t]{0.5\hsize}
	\centering
	\includegraphics[keepaspectratio,scale=0.21]{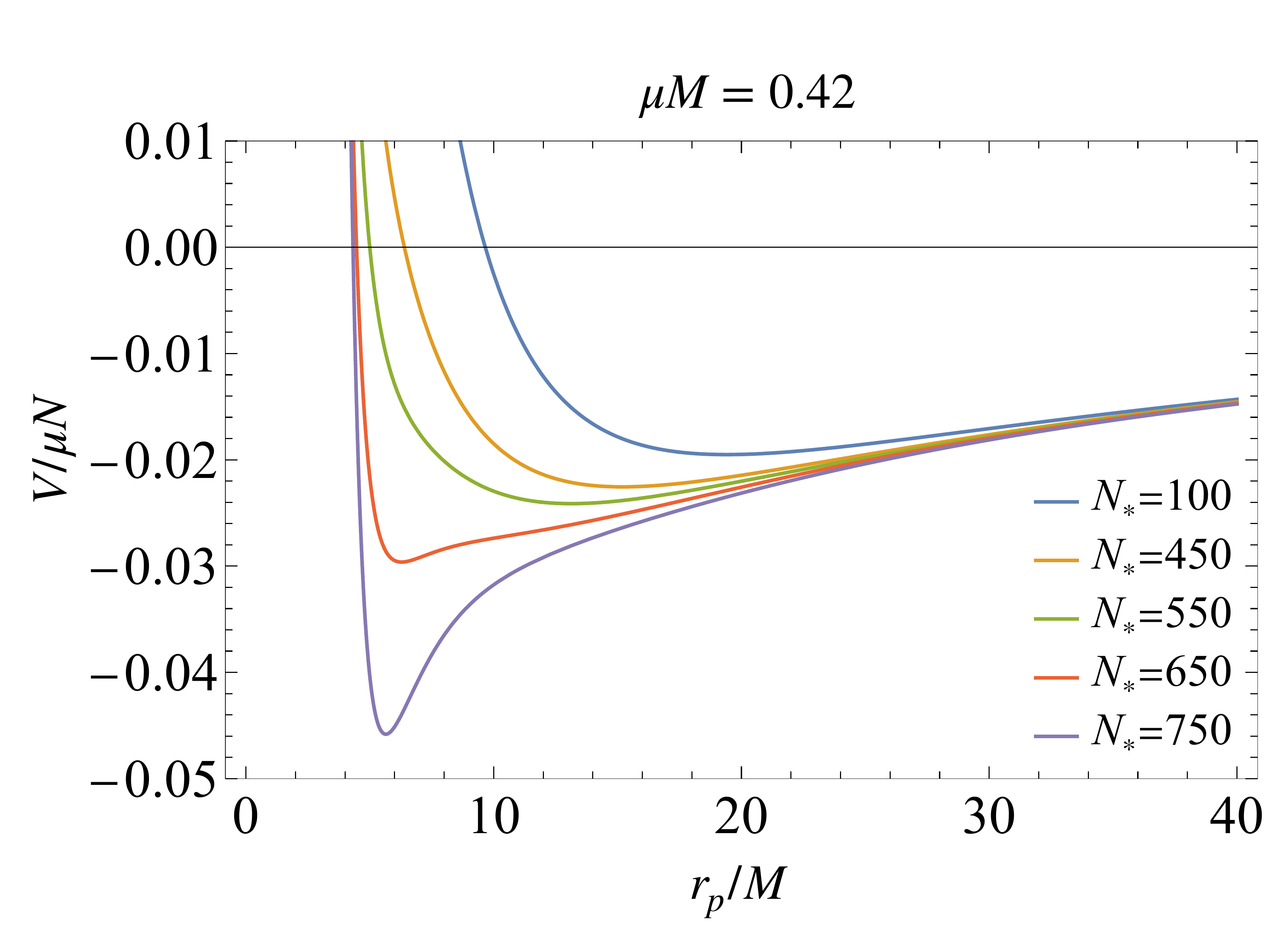}
\end{minipage} 
\end{tabular}
	\caption{Left(right) panel shows the dependence of the potential $V/\mu N|_{\sigma = \sigma_{eq}}$ for a fixed $N_*$ on $r_p$, for $\mu M =0.15$($0.42$). Each curve corresponds to a different value of $N_*$.}
	\label{fig:nonrelpot}
\end{figure}

\begin{figure}[t]
\begin{tabular}{cc}
\begin{minipage}[t]{0.5\hsize}
	\centering
	\includegraphics[keepaspectratio,scale=0.21]{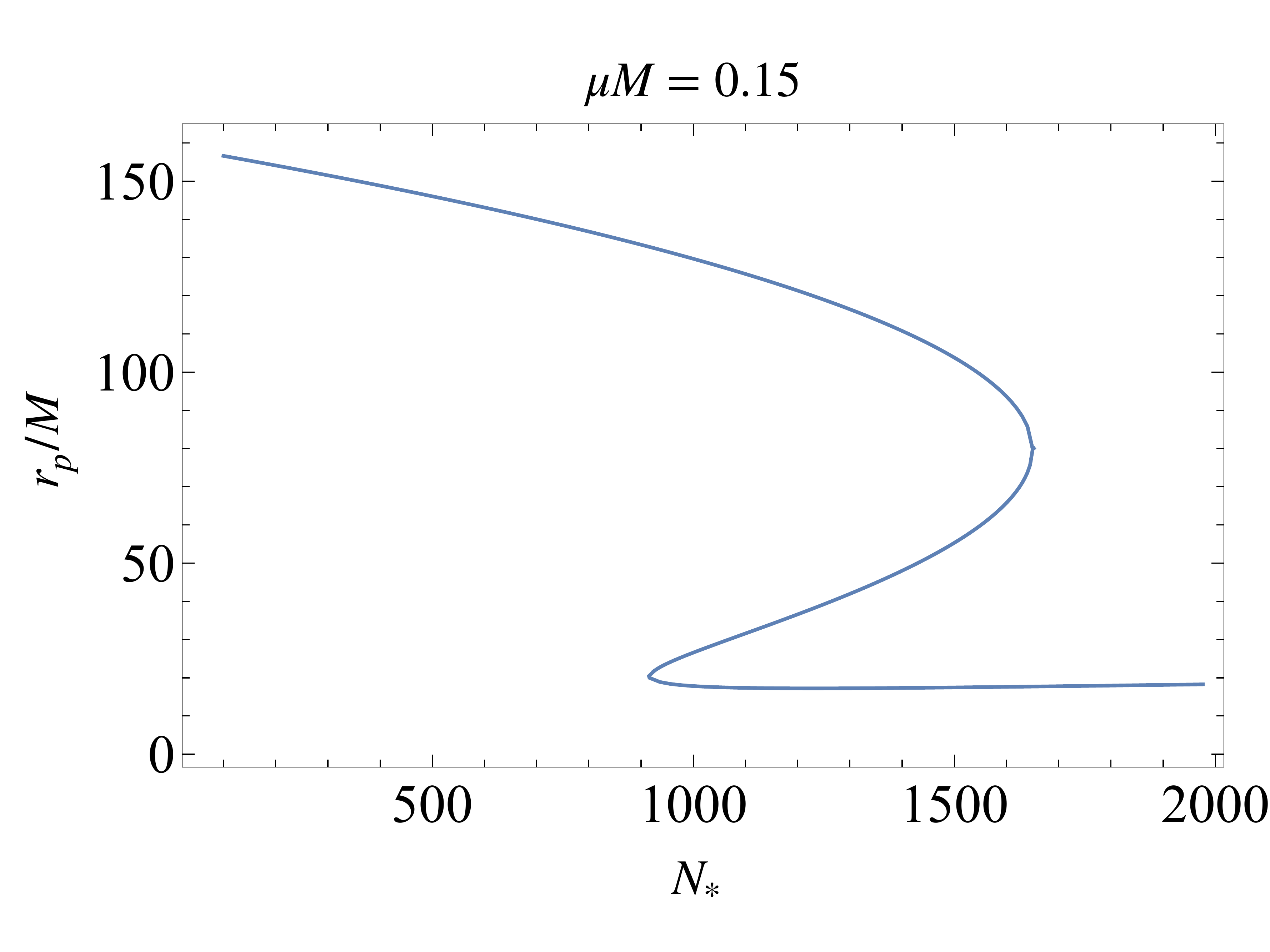}
\end{minipage}&
\begin{minipage}[t]{0.5\hsize}
	\centering
	\includegraphics[keepaspectratio,scale=0.21]{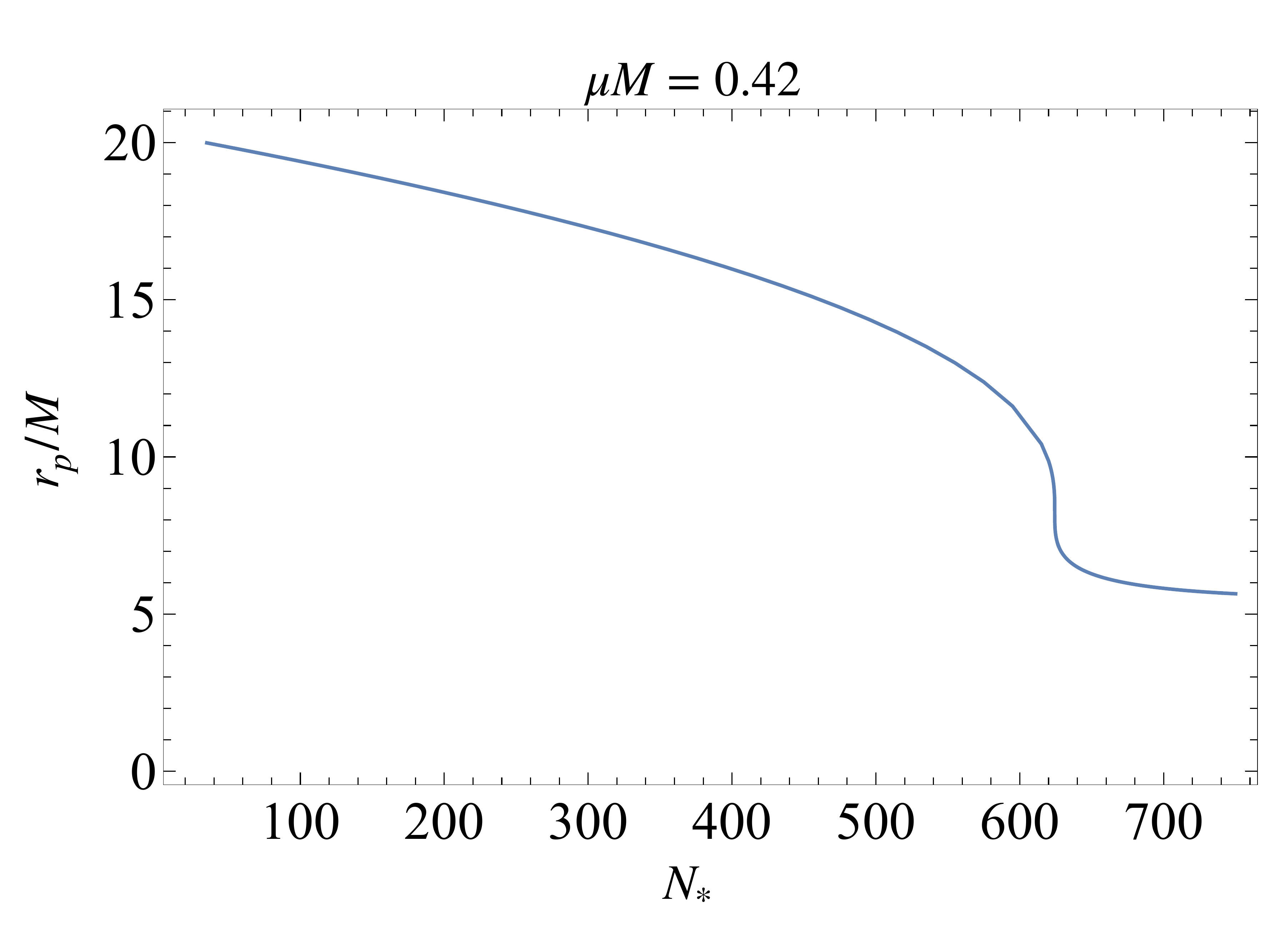}
\end{minipage} 
\end{tabular}
	\caption{Left(right) panel shows a value of $r_p$ at the extremum of the potential $V/N|_{\sigma = \sigma_{eq}}$ as a function of $N_*$ for $\mu M =0.15$($0.42$). 
	}
	\label{fig:extremrp}
\end{figure}

\begin{figure}[t]
\begin{tabular}{cc}
\begin{minipage}[t]{0.5\hsize}
	\centering
	\includegraphics[keepaspectratio,scale=0.21]{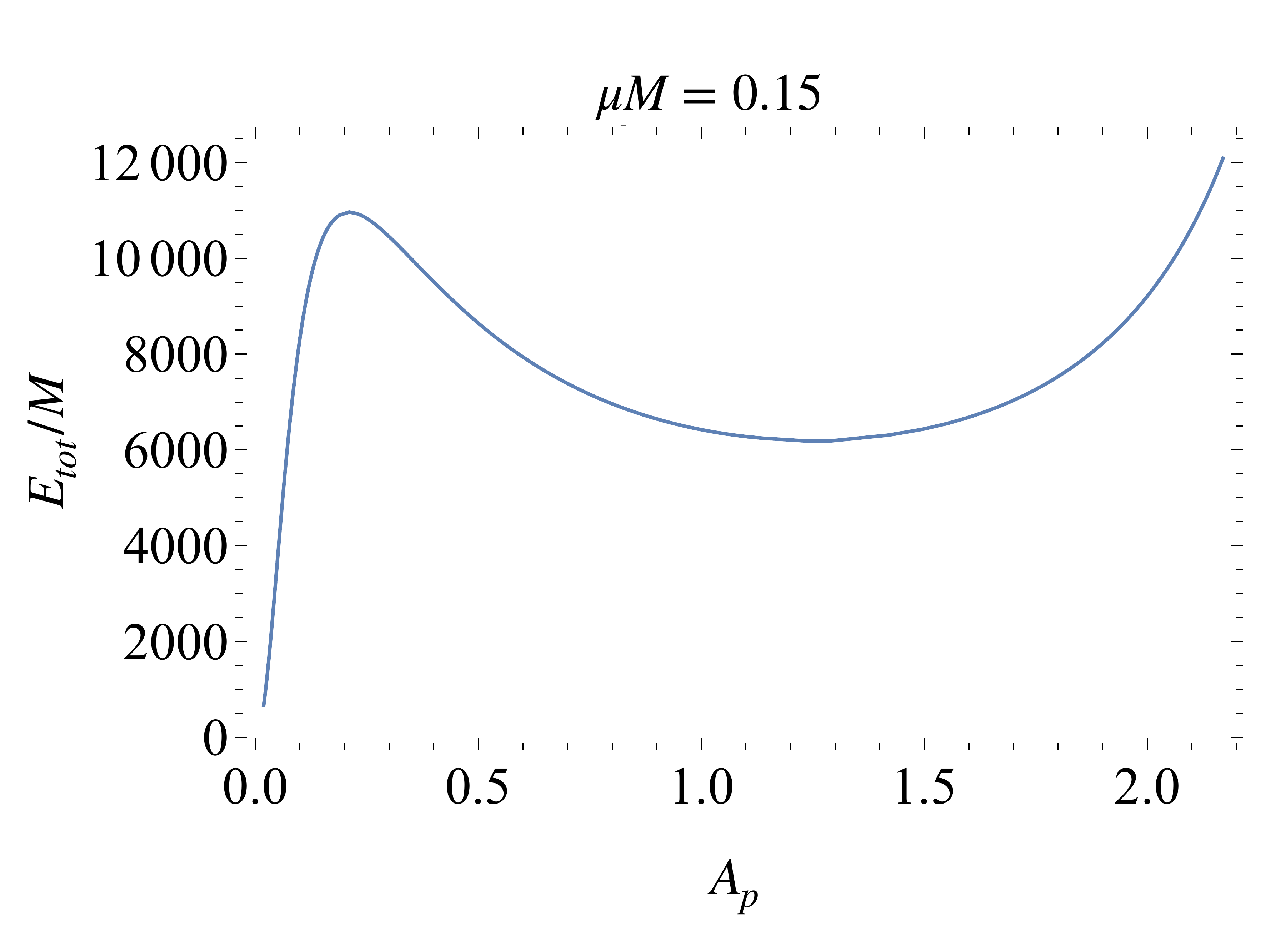}
\end{minipage}&
\begin{minipage}[t]{0.5\hsize}
	\centering
	\includegraphics[keepaspectratio,scale=0.21]{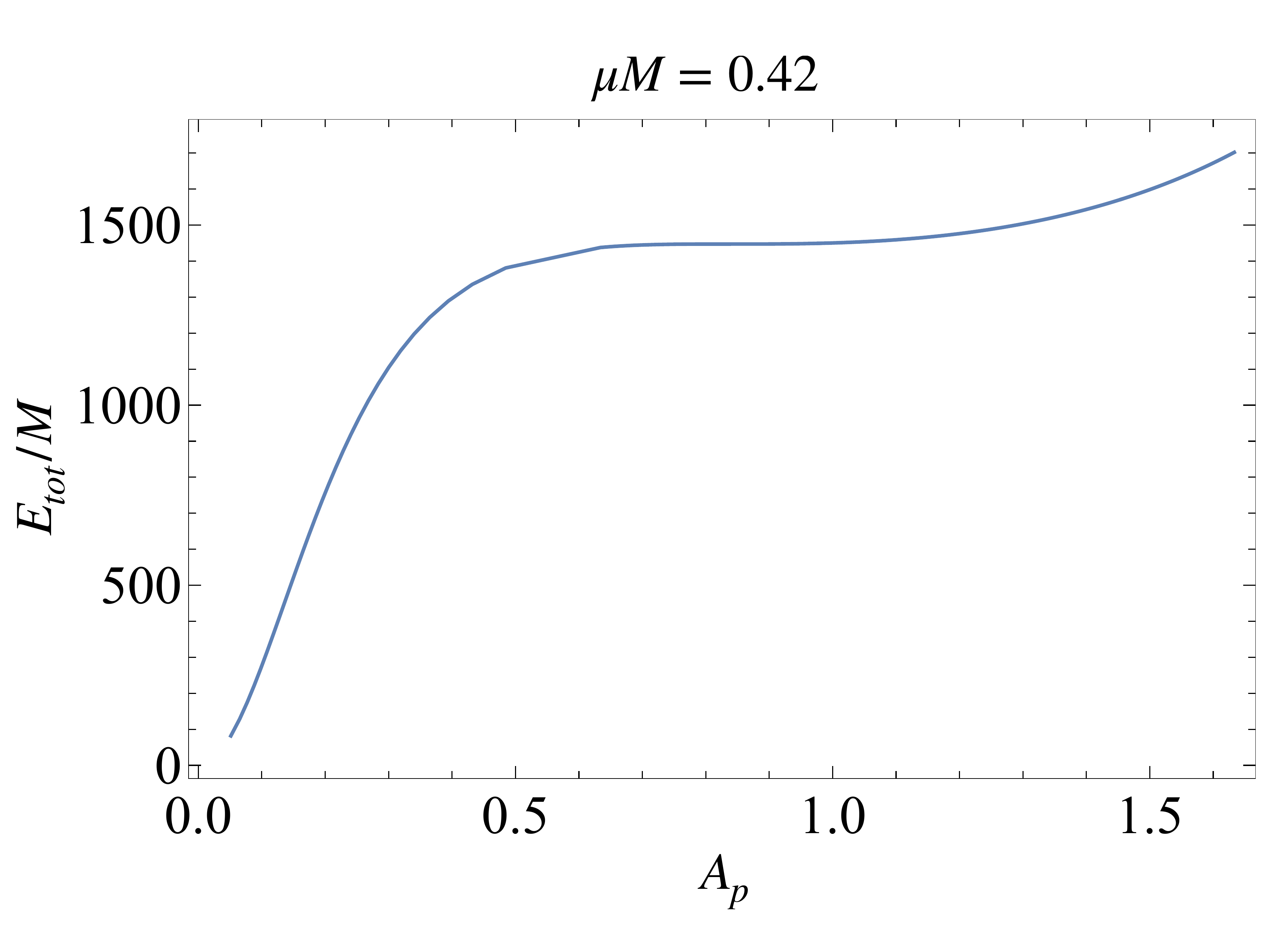}
\end{minipage} 
\end{tabular}
	\caption{Left(right) panel shows the dependence of the total energy $E_{\rm tot}$ of equilibrium configuration on the peak amplitude $A_p$ for $\mu M = 0.15$($0.42$).
	}
	\label{fig:toymodelenergy}
\end{figure}

The configuration of the cloud for a given particle number $N$ is determined by the extremum of the potential \eqref{eq:potfixN}, {\it i.e.}, 
\begin{align}
 \partial_{r_p} V = \partial_{\sigma} V = 0~.
\end{align}
Eliminating $N_*$ from these equations and solving for $\sigma$, we obtain
\begin{align}
    \sigma^2=\sigma^{2}_{eq} \equiv \frac{1}{6\mu^2 M}\left(-2 r_p + \mu^2 M r_p^2 + r_p \sqrt{4 + 2 \mu^2 M r_p + \mu^4 M^2 r_p^2}\right)~.
\end{align}
Here, $\sigma_{eq}$ denotes the radial width of the equilibrium configuration. After substituting $\sigma_{eq}$, we can regard the potential as a function of a single variable $r_p$. 
Figure \ref{fig:nonrelpot} shows the behavior of $V/N|_{\sigma=\sigma_{eq}}$ with $\mu M = 0.15$ (left) and $0.42$ (right), as a function of $N_*$. 
The value of $r_p$ at the extremum as a function of $N_*$ for various $\mu M$ is shown in Fig. \ref{fig:extremrp}. We also show how total energy of equilibrium configuration
\begin{align}
    E_{\rm tot} = \mu N + V|_{\sigma=\sigma_{eq}}
\end{align}
depends on $A_p$ in Fig. \ref{fig:toymodelenergy}.

We first study the case with $\mu M = 0.15$. 
For small $N_*(\lesssim 900)$, there is only one extremum made by the Newtonian potential and the angular momentum barrier. 
As cloud grows by the superradiant instability, $N_*$ becomes larger and $r_p$ decreases because of the attractive nature of the leading term $\propto \phi^4$ in the self-interaction. 
For $900 \lesssim N_* \lesssim 1650$, three extremum points, two stable and one unstable, appear. 
Outer stable point corresponds to the extremum in the small $N_*$ regime. 
Appearance of the inner stable point is due to the self-interaction. 
If we increase $N_*$ beyond $\sim 1650$, the outer stable point disappears and only the inner stable point remains. 
Therefore, the cloud residing at the outer stable point jumps to the inner stable point at $N_* \sim 1650$. This would be a clear indication of the onset of phase transition. 
Comparing the left panel of Fig. \ref{fig:toymodelenergy} to the right panel of Fig. \ref{fig:nonrelenergy}, the pattern of the instability is identical in both the numerical calculation and 
this toy model. 
Since the cloud jumps to the inner stable point when the potential barrier disappears, the phase transition can be violent and may cause an explosive phenomena, such as bosenova.
However, the dynamics of and the state after the phase transition cannot be studied by our method, and to clarify what really happens after the onset of the instability, dynamical simulations are necessary.

Now, we study the behavior of $\mu M = 0.42$ case. In this case there is only one extremum for any value of $N_*$. 
This is because the Newtonian potential becomes deeper and the radius where the gravitational force and the centrifugal force balance gets smaller, as $\mu M$ increases. Then, it becomes closer to the radius where the secondary minimum due to the self-interaction appears, and finally the range of amplitude in which two local minimums coexist disappears.
Since no phase transition occurs, the energy flux to infinity, which is not included in this toy model, balances with the superradiant growth at some $N_*$, and the growth terminates there. This agrees with our numerical calculation in sec. \ref{sec:4.1}.

\begin{figure}[t]
\begin{tabular}{cc}
\begin{minipage}[t]{0.5\hsize}
	\centering
	\includegraphics[keepaspectratio,scale=0.21]{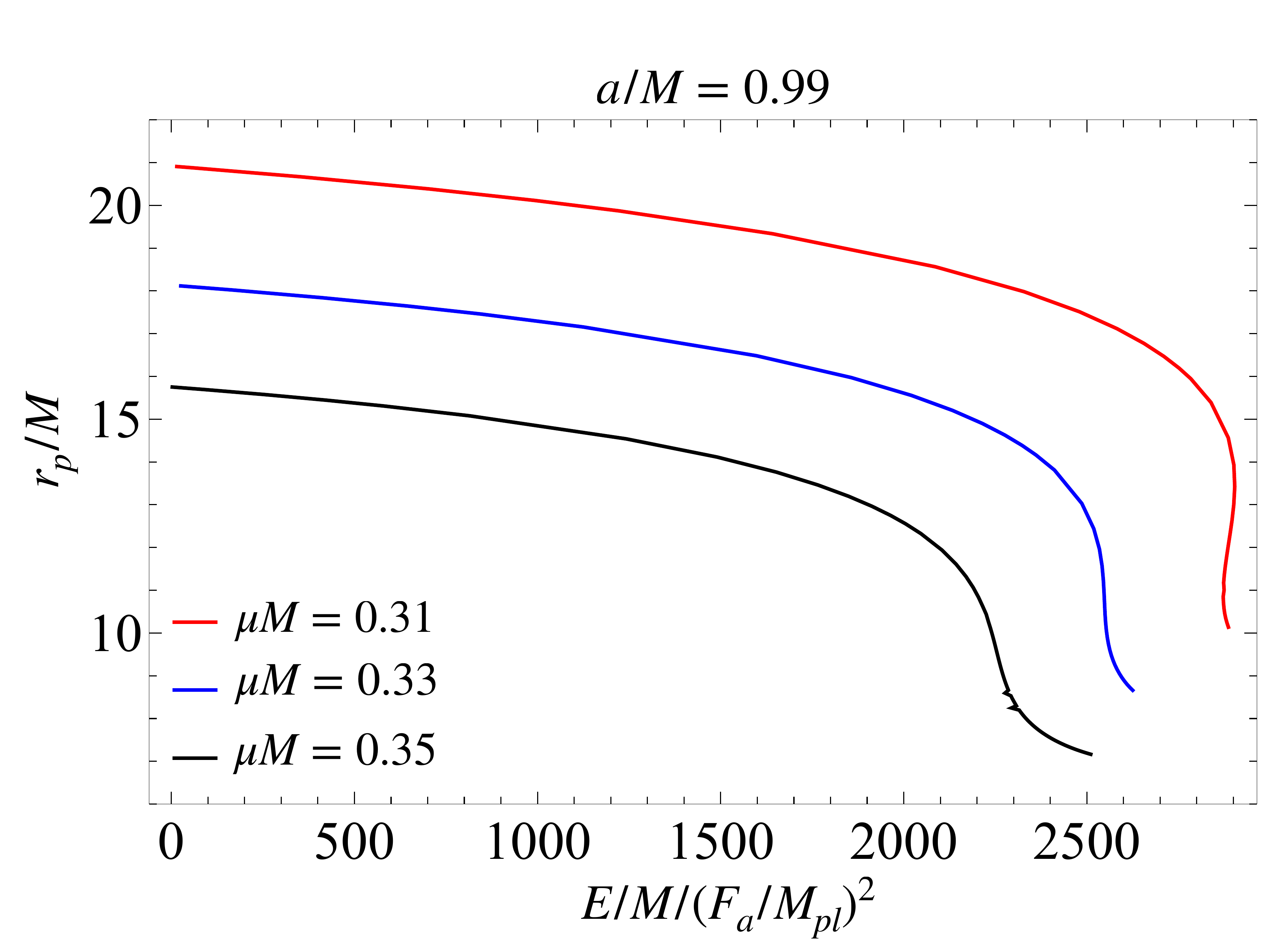}
\end{minipage}&
\begin{minipage}[t]{0.5\hsize}
	\centering
	\includegraphics[keepaspectratio,scale=0.21]{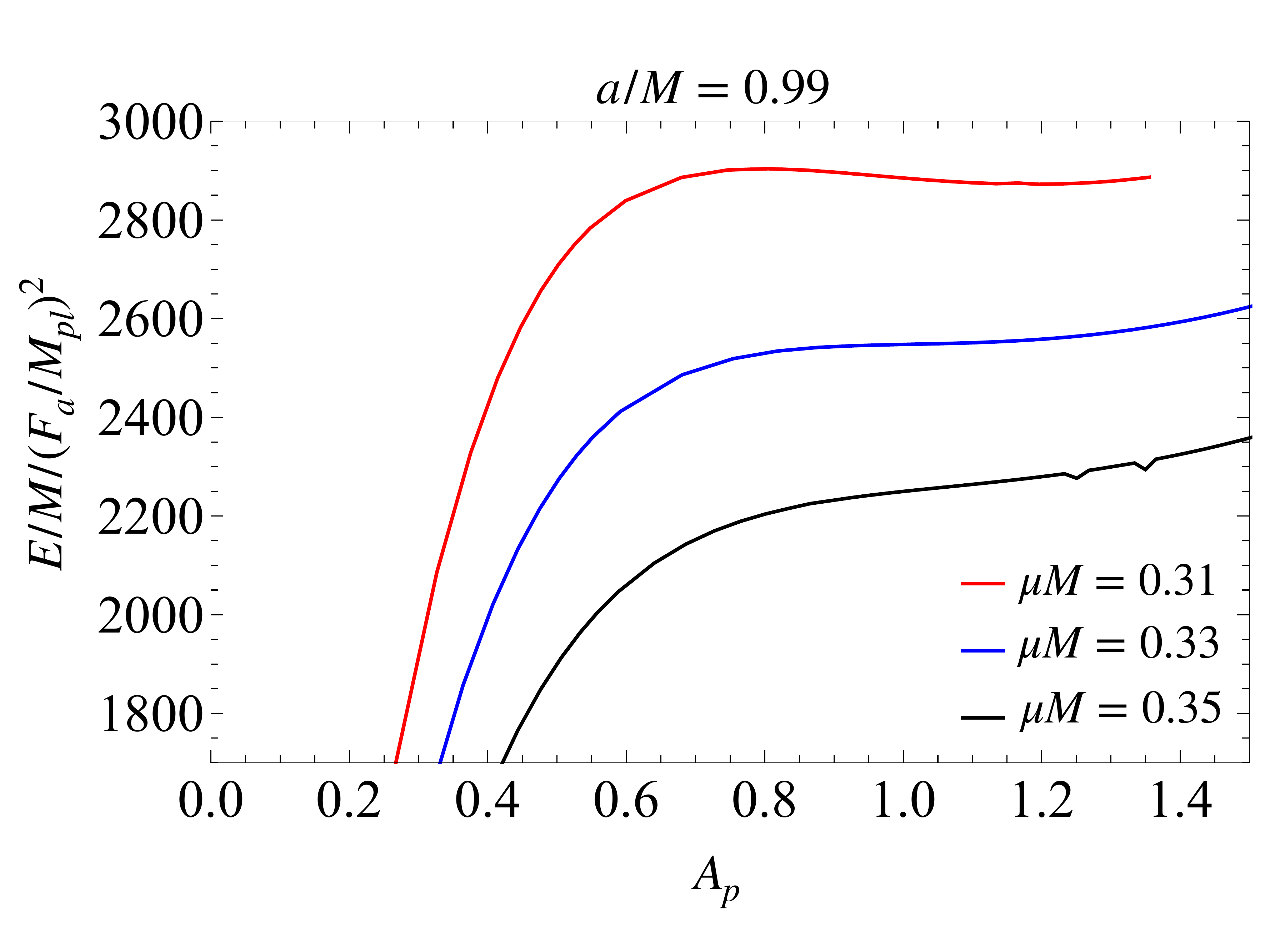}
\end{minipage} 
\end{tabular}
	\caption{(Left) Red, blue, and black curves show the position of the peak of the fundamental mode $\tilde{R}_{11}$ as a function of the energy $E$ for $\mu M = 0.31,0.33,$ and $0.35$, respectively. The spin of central BH is fixed to $a/M=0.99$. Different configurations with a given $E$ can exist only for $\mu M = 0.31$ case. (Right) Red, blue, and black curves show the energy of the configuration at the extremum as a function of the peak amplitude of the fundamental mode $\tilde{R}_{11}$ for $\mu M = 0.31,0.33,$ and $0.35$, respectively. A local maximum in this plot appears only for $\mu M =0.31$ case.}
	\label{fig:critical}
\end{figure}

Our toy model suggests that there exists a critical gravitational coupling $\mu_c M$, above which no phase transition occurs. 
In the case just below this critical value, the phase transition 
might be very mild, even though it is the first order phase transition. 

To determine the critical value, $\mu_c M$, we further calculate the evolution of axion clouds for various values of $\mu M$ with the method presented in sec. \ref{sec:3}. 
In Fig \ref{fig:critical} we give the same plots as Figs. \ref{fig:extremrp} (left panel) and \ref{fig:toymodelenergy} (right panel) but with $\mu M = 0.31, 0.33,$ and $0.35$ for $a/M=0.99$. 
From these figures, we find that the critical value is around $\mu_c M \sim 0.32$.

\section{Summary and Discussion}\label{sec:6}

In this paper, we numerically examined the effect of self-interaction on the evolution of an axion cloud, under the assumption that the evolution is adiabatic. 
Our main focus is to establish a method to track a consistent evolution of clouds, 
starting with a very small amplitude to a large amplitude such that we can transfer the data of the field configuration to a fully dynamical simulation as an appropriate initial data. 
In this paper we have restricted our attention to the case in which only the $l=1$, $m=1$ fundamental superradiant mode is initially occupied by the axion cloud. 
We first investigated the case with the axion mass  $\mu M = 0.42$ and the central BH spin $a/M = 0.99$, which realizes the growth rate around the maximum. We found that the cloud with an attractive self-interation $\propto \phi^4$ only becomes unstable at a certain energy. On the other hand, we found that, when we include the higher order terms in the cosine-type potential, no instabilities occur and cloud settles into a quasi-stationary state, where the energy dissipation to infinity balances with the energy gain due to the superradiance. 
This is because the cloud remains stable throughout the evolution owing to the repulsive force from the higher order terms in the cosine-type potential, and the energy dissipation to infinity eventually becomes sufficiently large as the amplitude of the cloud increases.

Moreover, we investigated how the evolution of clouds depends on the parameters $(\mu M, a/M)$. 
We found that for a large $\mu M$, clouds evolve into a quasi-stationary state as we obtained in $\mu M = 0.42$ case. 
As we decrease the value $\mu M$, there appears a critical value at which the cloud becomes unstable. From our calculation, the instability occurs when $E/M \sim 3\times 10^2/(G\mu M)^2(F_a/M_{pl})^2$ \footnote{Here, we assumed that the energy at the onset of the instability scales on gravitational coupling as $\propto(G\mu M)^{-2}$, which is motivated by the estimation of the energy when the bosenova happens using the non-relativistic approximation \cite{Arvanitaki:2010sy,Baryakhtar:2020gao}}. 
Also, we found that the BH spin does not have any significant influence 
on whether or not the instability occurs. 
The main role of the BH spin is to control the existence of the superradiance and the superradiant instability time scale.

Next, we constructed a toy model which describes the behavior of the cloud found by our numerical calculation. 
We approximate the cloud to be a Gaussian distribution characterized by only three parameters, the peak amplitude $A_p$, the radial position of the peak $r_p$, and the radial extension of the cloud $\sigma$. 
For a large $\mu M$, there exists only one equilibrium configuration throughout the evolution. 
This means that the cloud is stable throughout the evolution. 
By contrast, three distinct equilibrium configurations, two stable and one unstable, can exist for a small $\mu M$, in a certain energy range. 
Therefore, a sudden change between two stable equilibriums, {\it i.e.}, the first order phase transition, can happen as one increases the energy. We interpret this phase transition in our toy model as the onset of instability suggested from our numerical calculation by the appearance of the peak in the cloud energy as a function of the cloud amplitude. 
With the aid of this toy model, we can confidently conclude that no occurrence of bosenova is expected for a large $\mu M$. 
Further numerical calculation showed that the critical gravitational coupling 
is around $\mu_c M \sim 0.32$.

Since our calculation is relying on the adiabatic approximation, we cannot tell what happens after the onset of the instability. 
To clarify the fate of the instability, dynamical simulations are necessary. 
Moreover, we have not confirmed the dynamical stability of the quasi-stationary configurations for a large $\mu M$. Thus, it might be too early to conclude that the cloud evolves to a quasi-stationary state for a large $\mu M$.
These issues can also be clarified by dynamical simulations.

It should be noted that we have ignored saturation mechanisms other than the axion emission to infinity, such as the spin-down of the BH \cite{Arvanitaki:2010sy} and the energy dissipation due to the existence of the multiple superradiant modes \cite{Baryakhtar:2020gao}. 
If another saturation mechanism works before the onset of instability, 
we cannot expect an explosive phenomenon to happen. We can assess whether the spin-down of the BH can be effective or not by looking at the angular momentum of the cloud. Let us first analyze with unstable case, {\it i.e.} $\mu M \lesssim 0.32$. For $\mu M = 0.29$ case, the angular momentum of the cloud at around the maximum of the energy is 
\begin{align}\label{eq:angm029}
J_{cl}/J_{BH} \sim 1.1\times 10^4(F_a/M_{pl})^2(1/(a/M))~.
\end{align}
This is around $10^{-2}$ of that of central BH for the GUT scale decay constant $F_{a}/M_{pl} \sim 10^{-3}$, which is small enough to neglect the spin-down of the BH. For a smaller decay constant, this fraction is even smaller. Therefore, the saturation of the superradiance condition \eqref{eq:SRcond} due to the spin-down of the central BH is not very likely before the instability sets in. 

On the other hand, for $\mu M = 0.15$, the angular momentum of the cloud when the instability sets in is roughly given by  
\begin{align}
J_{cl}/J_{BH} \sim 9\times 10^4(F_a/M_{pl})^2(1/(a/M))~. 
\end{align}
Thus, the change of the BH spin is around $\Delta (a/M) \sim 0.1$ for the GUT scale decay constant, which means that the spin-down of the BH cannot be always neglected for a small $\mu M$. 
When the initial BH spin is close to the value for the saturation of the superradiance condition ({\it e.g.} $a/M \sim 0.55$ for $\mu M = 0.15$ case), the growth of cloud due to superradiance would terminate before the instability sets in. 
By contrast, for a large spin ($a/M \sim 0.9$) which gives a larger growth rate, the saturation of superradiance condition will not occur before the instability sets in, even if we consider the spin down of the BH. To summarize, as long as one considers a large BH spin $a/M \gtrsim 0.9$, the spin down of the BH does not prevent the ignition of the instability.

For the saturation case, $\mu M \gtrsim 0.32$, the evolution of the cloud is not so affected as in the $\mu M = 0.29$ case. The angular momentum of the non-linear quasi-stationary configuration is smaller than Eq. \eqref{eq:angm029} (for example, $J_{cl}/J_{BH} \sim 4.5\times 10^3 (F_a/M_{pl})^2 (0.99/(a/M))$ for $\mu M = 0.42$). Therefore, BH spin-down can be neglected for the GUT scale decay constant before the saturation due to self-interaction occurs. Thus, the cloud first settles to a non-linear configuration as shown in Fig. \ref{fig:fieldconfig042}. Then, the angular momentum of the central BH is extracted by the superradiance in a longer time scale. As the BH spin gets smaller, the energy flux from the horizon gets smaller. To satisfy the balance of the total energy flow, the cloud expands to reduce the energy flux to infinity. In this manner, the BH spin-down proceeds as long as the superradiance condition is satisfied. During the process of the BH spin-down, the cloud maintains the quasi-stationary configuration. After the saturation of the superradiance condition, the cloud gradually dissipates the energy to infinity by the self-interaction. To qualitatively predicts the final value of the BH spin, further numerical calculation which takes into account the time evolution of BH mass and spin is necessary.

The effect of multiple modes also cannot be studied within our formalism as it is, since the presence of the second superradiant mode breaks the helical symmetry. 
Since perturbative calculation suggests that the dissipation due to multiple modes works efficiently in the strongly nonlinear regime (especially for the relativistic cloud)\cite{Omiya:2020vji}, precise calculation without relying on perturbative analysis is needed to tell whether dissipation may terminate the growth before the instability occurs or not.
In the preceding studies, for example, the deformation of the cloud due to the self-interaction has not been taken into account.
Since the strengths of the mode coupling between different modes are determined by the size of overlap 
between the modes, the deformation of the cloud might have a significant impact on the rate of dissipation. Related to this point, the gravitational wave emission from the cloud has to be reinvestigated. 
Because the radial extension of the cloud shrinks owing to the self-interaction, 
it becomes comparable to the wavelength of relevant gravitational waves. 
Thus, the energy flux carried out by gravitational waves can be enhanced compared with the naive estimate based on the linearized model. 
These points would be further discussed in the future work.

\section*{Acknowledgements}
We thank Hirotaka Yoshino for his helpful comments. This work is supported by JSPS Grant-in-Aid for Scientific Research JP17H06358 (and also JP17H06357), as a part of the innovative research area, ``Gravitational wave physics and astronomy: Genesis'', and also by JP20K03928. 
T. Takahashi is supported by "the establishment of university fellowships towards the creation of science technology innovation". 

\appendix

\section{Details of the numerical calculation}\label{app:A}

Here, we briefly summarize our calculation scheme to solve Eq. \eqref{eq:Rlmeq} under the boundary conditions \eqref{eq:BClin} and \eqref{eq:BClinH}. We truncate the infinite summation in Eq. \eqref{eq:adiabaticansatz} at $l_{\rm max} = 5, n_{\rm max} = 5$. We confirm that truncating $l$ and $n$ at these values does not change the results much, as presented in appendix \ref{app:B}. Since we start with $l_0 =1,m_0=1$ and the potentials (Eqs. \eqref{eq:phi4} - \eqref{eq:cos}) are even functions of $\phi$, we only need to consider modes with odd $l,m$. Thus, only modes with $(l,m) = (1,1),(3,1),(5,1),(3,3),(5,3),(5,5)$ appear in our calculation. 

Our task is to determine the frequency $\omega_0$ and the amplitudes of modes at the horizon and at a large radius. For a given amplitude of the fundamental mode at a large $r$, we determine remaining 12 complex parameters ($\omega_0$ and the remaining complex amplitudes) by matching the mode functions, obtained by solving the equations from the $r_* = r_{\rm min}$ with the boundary condition \eqref{eq:BClinH} and those from $r_* = r_{\rm max}$ with the boundary condition \eqref{eq:BClin}, at $r_* = r_{\rm match}$. In our calculation we take $r_{\rm min} = -100M, r_{\rm max} = 100M,$ and $r_{\rm match} = 5M$. We obtain the 12 parameters by starting with a small amplitude $(A_0 = 10^{-3})$ and gradually increase the amplitude. To obtain a new solution, we solved a set of linearized equations expanded around an approximate solution guessed by the extrapolation from the previous steps. Then, to keep the precision of the solution, we iteratively solve a set of linearized equations expanded around the improved approximate solution, until the convergence is confirmed.  
When we perform the integration of nonlinear terms over the angular coordinates in Eq. \eqref{eq:Rlmeq}, we use the orthogonal collocation method. We discretize the $\theta$ direction by $16$ points and $\varphi$ direction by $31$ points.

\section{Justification of the truncation of $l$ and $n$}\label{app:B}

\begin{figure}[t]
\begin{tabular}{cc}
\begin{minipage}[t]{0.5\hsize}
	\centering
	\includegraphics[keepaspectratio,scale=0.2]{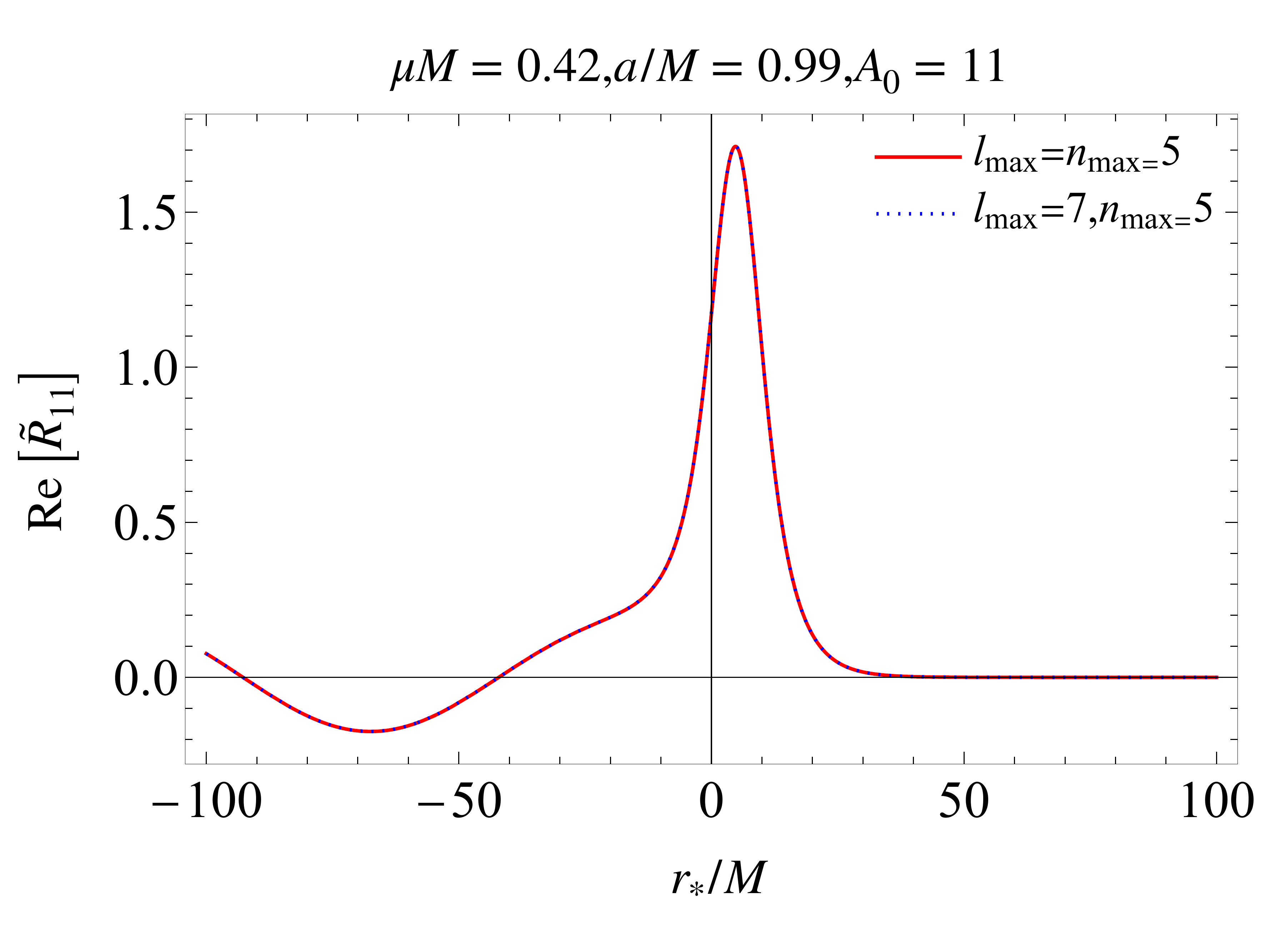}
\end{minipage} &
\begin{minipage}[t]{0.5\hsize}
	\centering
	\includegraphics[keepaspectratio,scale=0.2]{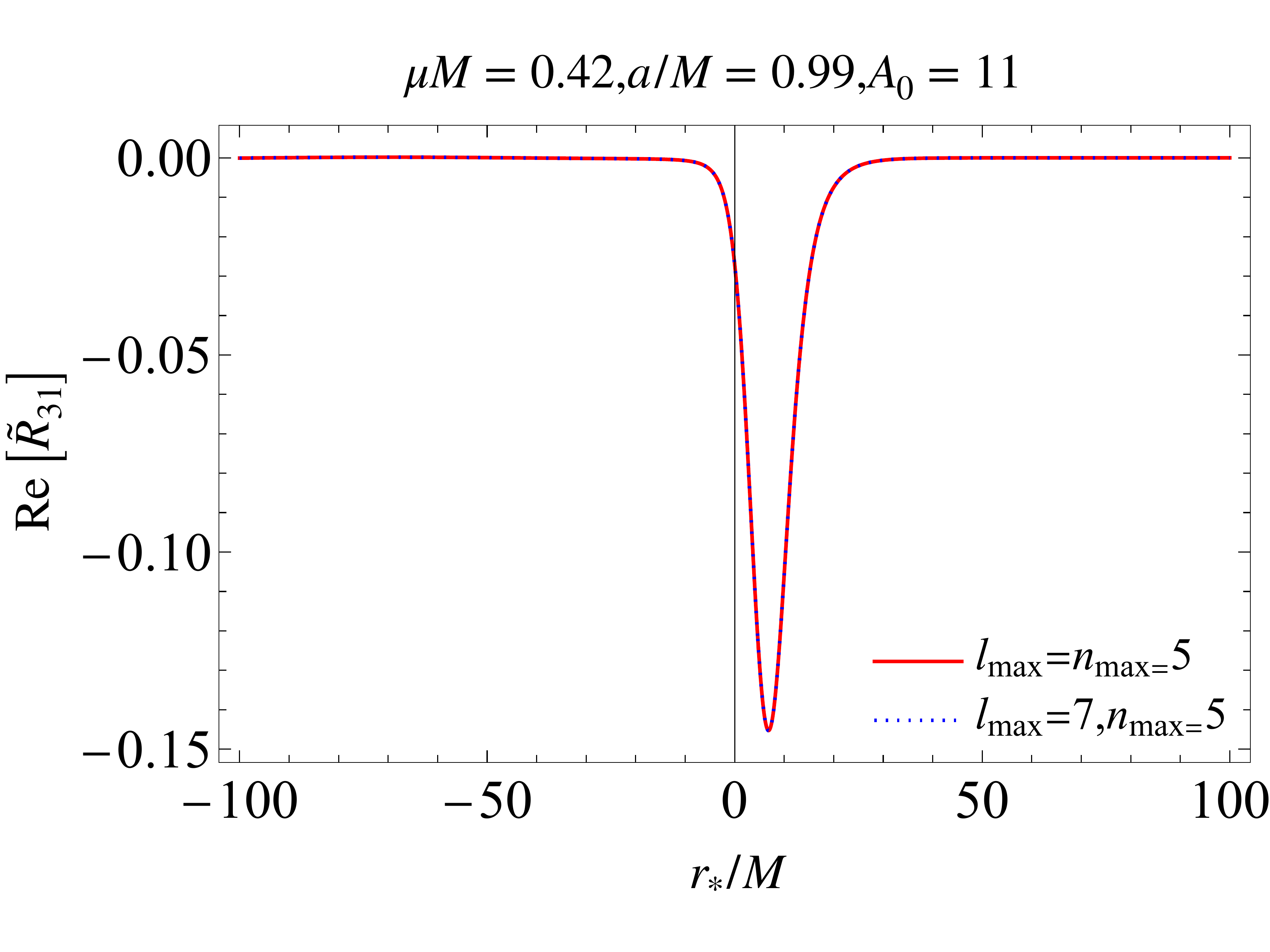}
\end{minipage}
\end{tabular}
\begin{tabular}{cc}
\begin{minipage}[t]{0.5\hsize}
	\centering
	\includegraphics[keepaspectratio,scale=0.2]{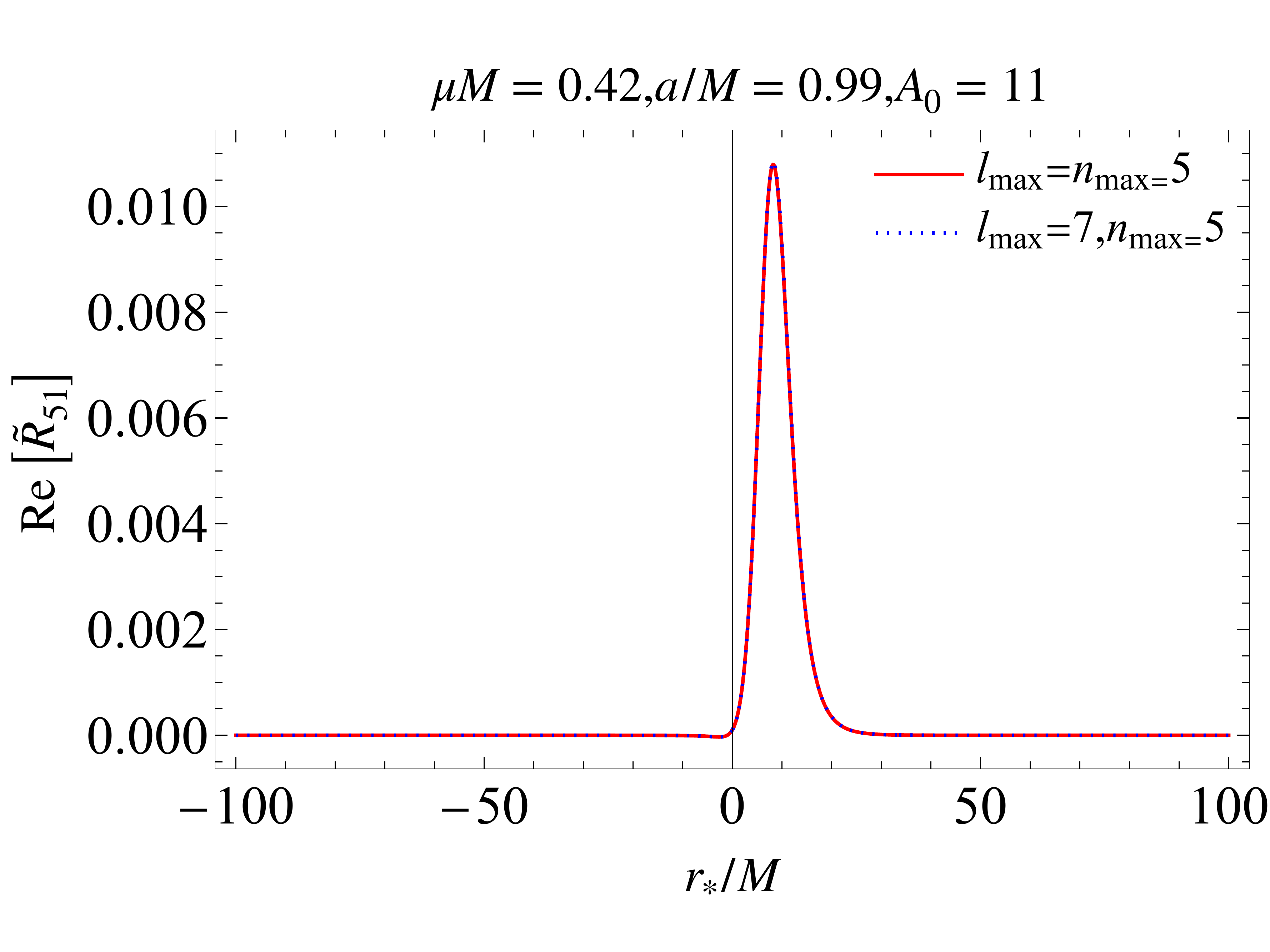}
\end{minipage} &
\begin{minipage}[t]{0.5\hsize}
	\centering
	\includegraphics[keepaspectratio,scale=0.2]{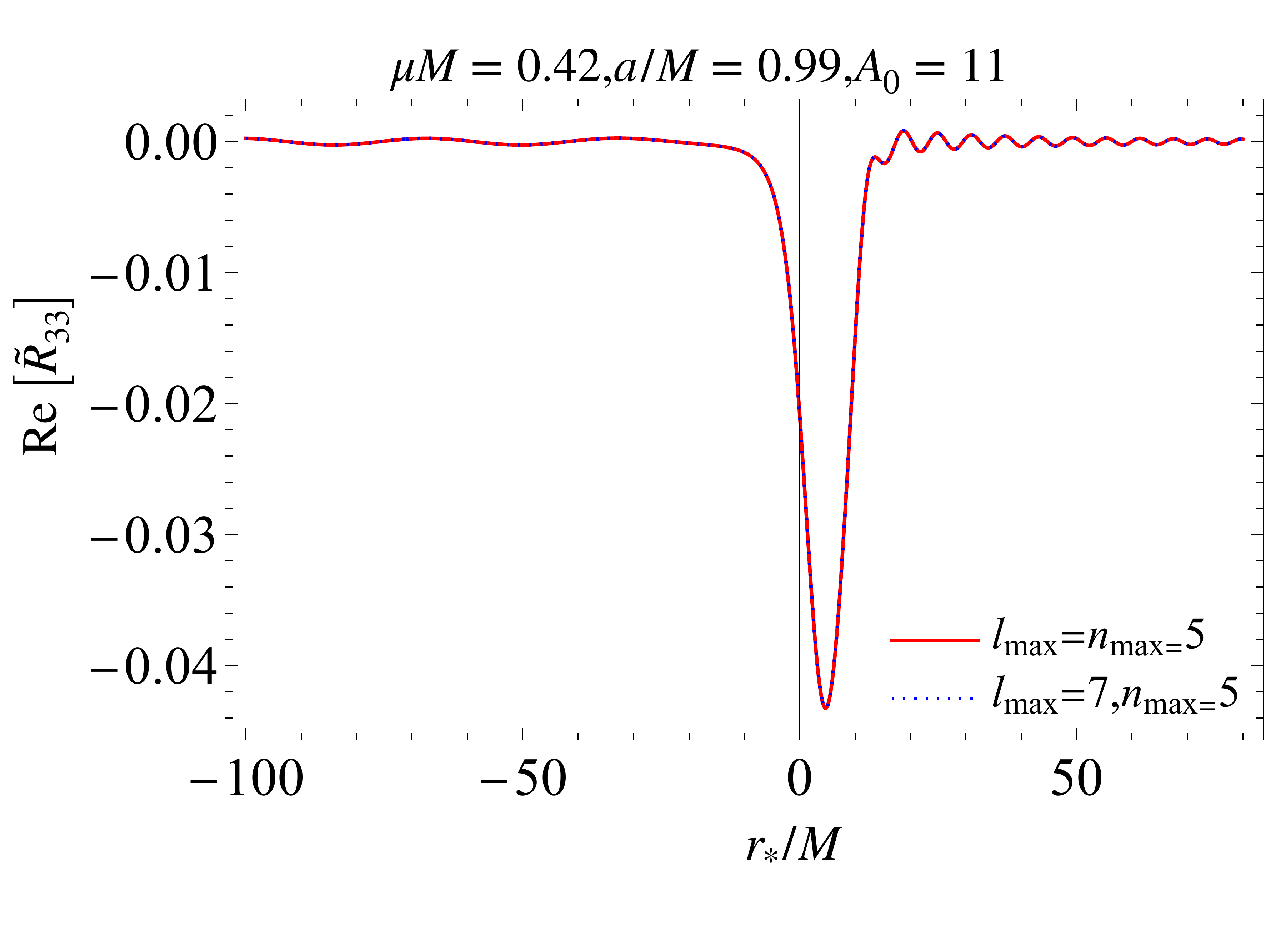}
\end{minipage}
\end{tabular}
\begin{tabular}{cc}
\begin{minipage}[t]{0.5\hsize}
	\centering
	\includegraphics[keepaspectratio,scale=0.2]{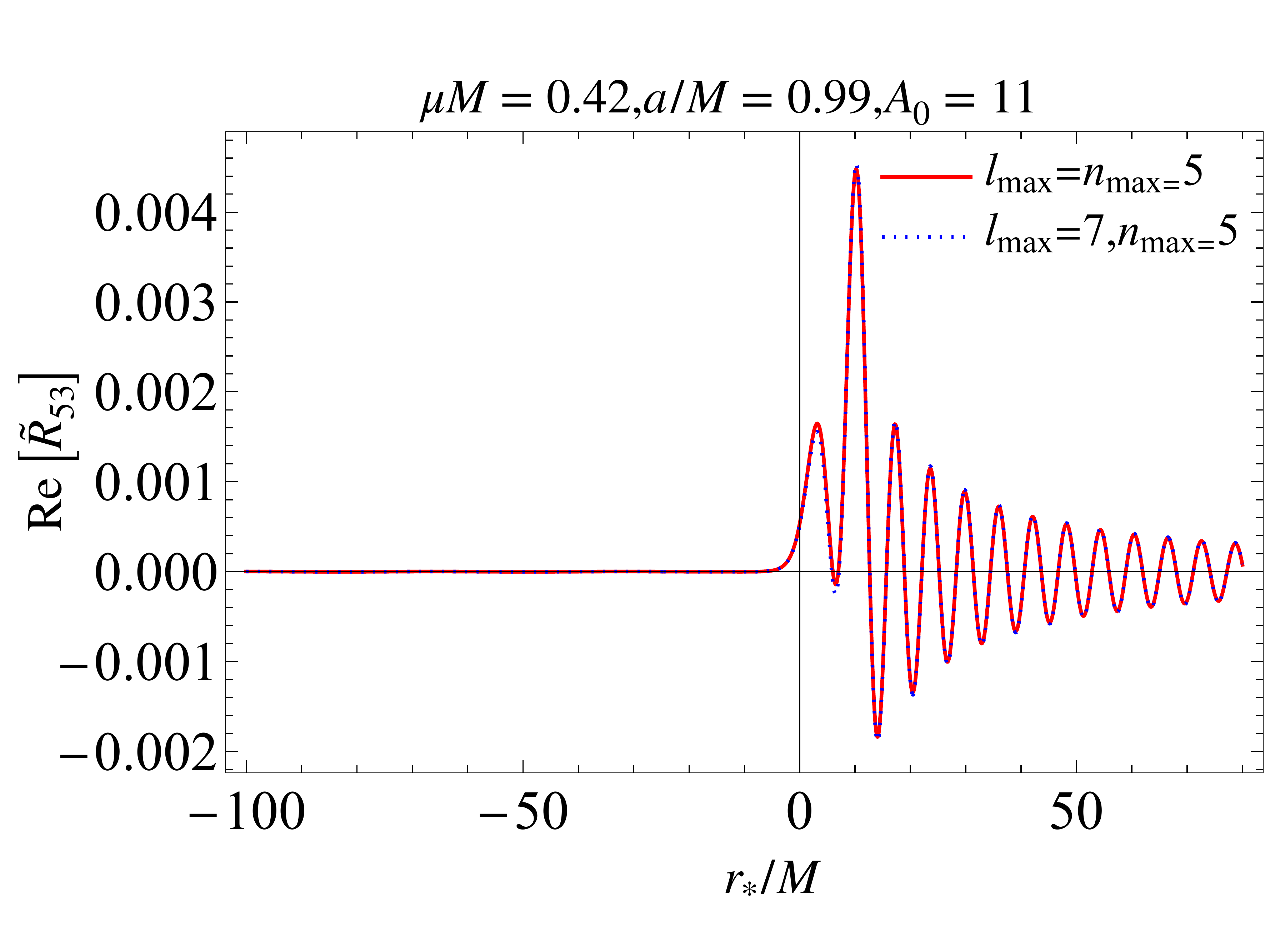}
\end{minipage} &
\begin{minipage}[t]{0.5\hsize}
	\centering
	\includegraphics[keepaspectratio,scale=0.2]{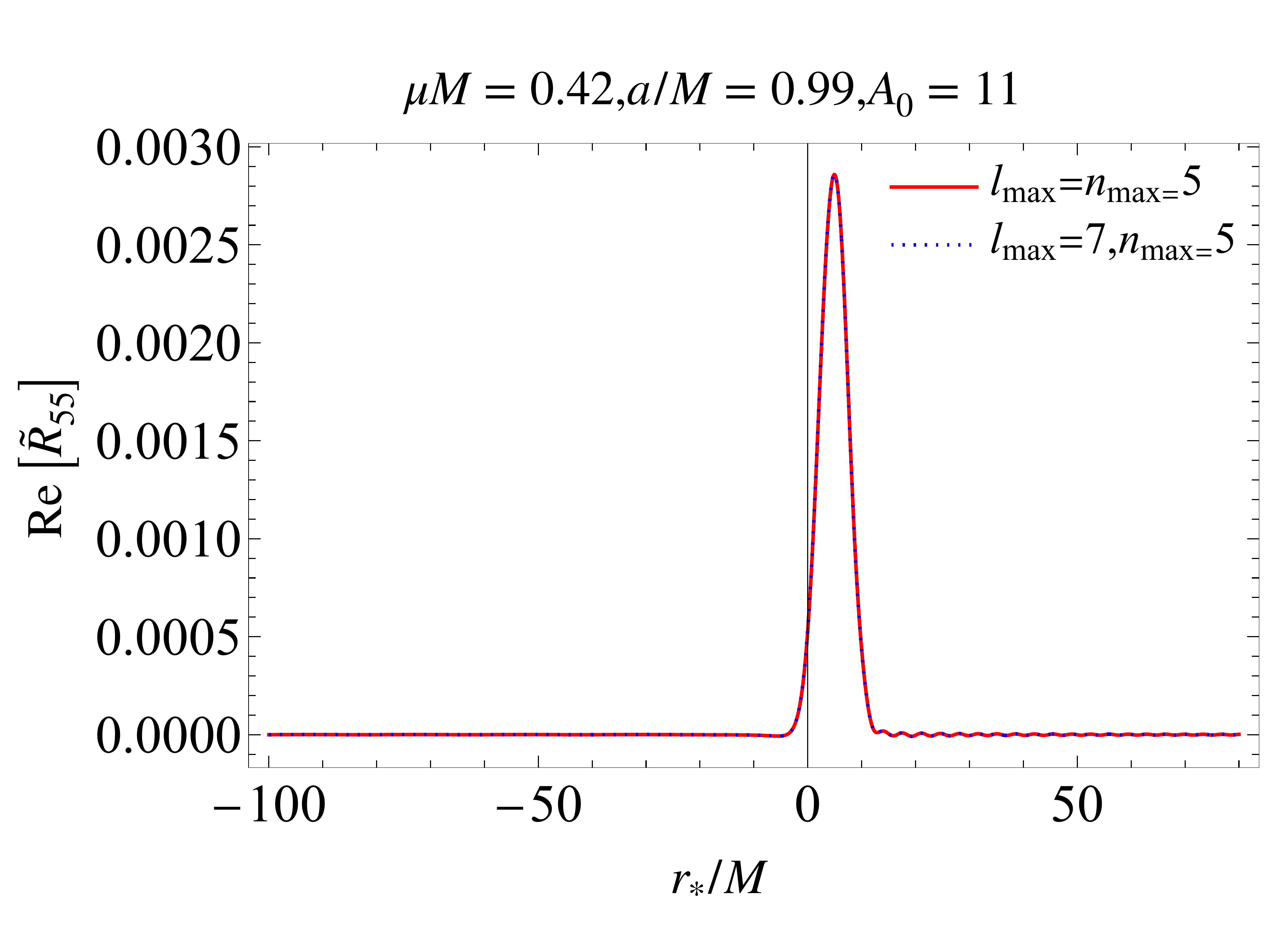}
\end{minipage}
\end{tabular}
	\caption{Each panel shows the real part of mode function $\tilde{R}_{lm}$ at amplitude $A_0 = 11$. From the top left to the bottom right,  $\tilde{R}_{11},\tilde{R}_{31},\tilde{R}_{51},\tilde{R}_{33},\tilde{R}_{53}$ and $\tilde{R}_{55}$ are plotted, respectively. The red solid curve is calculated with six modes $(l,m) = (1,1),(3,1),(5,1),(3,3),(5,3),$ and $(5,5)$, while the blue dashed curve is calculated by adding two more modes $(l,m) = (7,1)$ and $(7,3)$.}
	\label{fig:modefnc}
\end{figure}

In this appendix, we show the result of the same calculation for $\mu M = 0.42, a/M = 0.99$, but including the additional $(l,m) = (7,1),$ and $(7,3)$ modes. These two modes are enough to justify the truncation at $l_{\rm max} =5$ and $n_{\rm max}=5$, since modes with $m \geq 5$ are suppressed compared with $m = 1,3$ modes (see Fig. \ref{fig:modefnc} for example). Here, we fix the potential of the axion to be \eqref{eq:cospotential}.

\begin{figure}[t]
	\centering
	\includegraphics[keepaspectratio,scale=0.3]{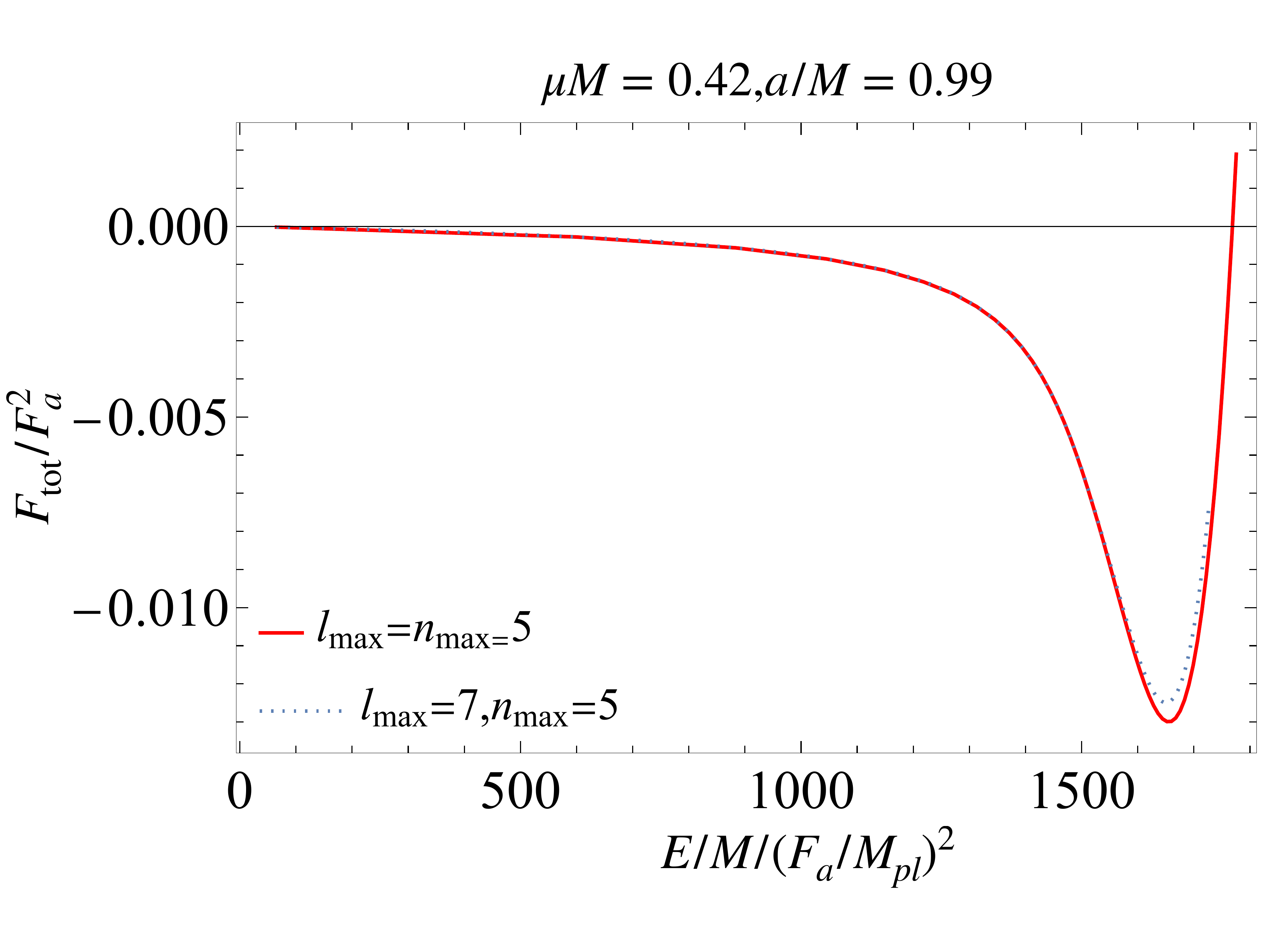}
	\caption{Dependence of the total flux $F_{\rm tot}(A_0)$ on the energy $E$. The red solid curve is the same as the blue solid curve in the right panel of Fig. \ref{fig:energyandflux}. The blue dotted curve corresponds to the calculation with additional higher multipole modes $(l,m) = (7,1)$ and $(7,3)$.}
	\label{fig:energyandfluxhighl}
\end{figure}

\begin{figure}[t]
	\centering
	\includegraphics[keepaspectratio,scale=0.3]{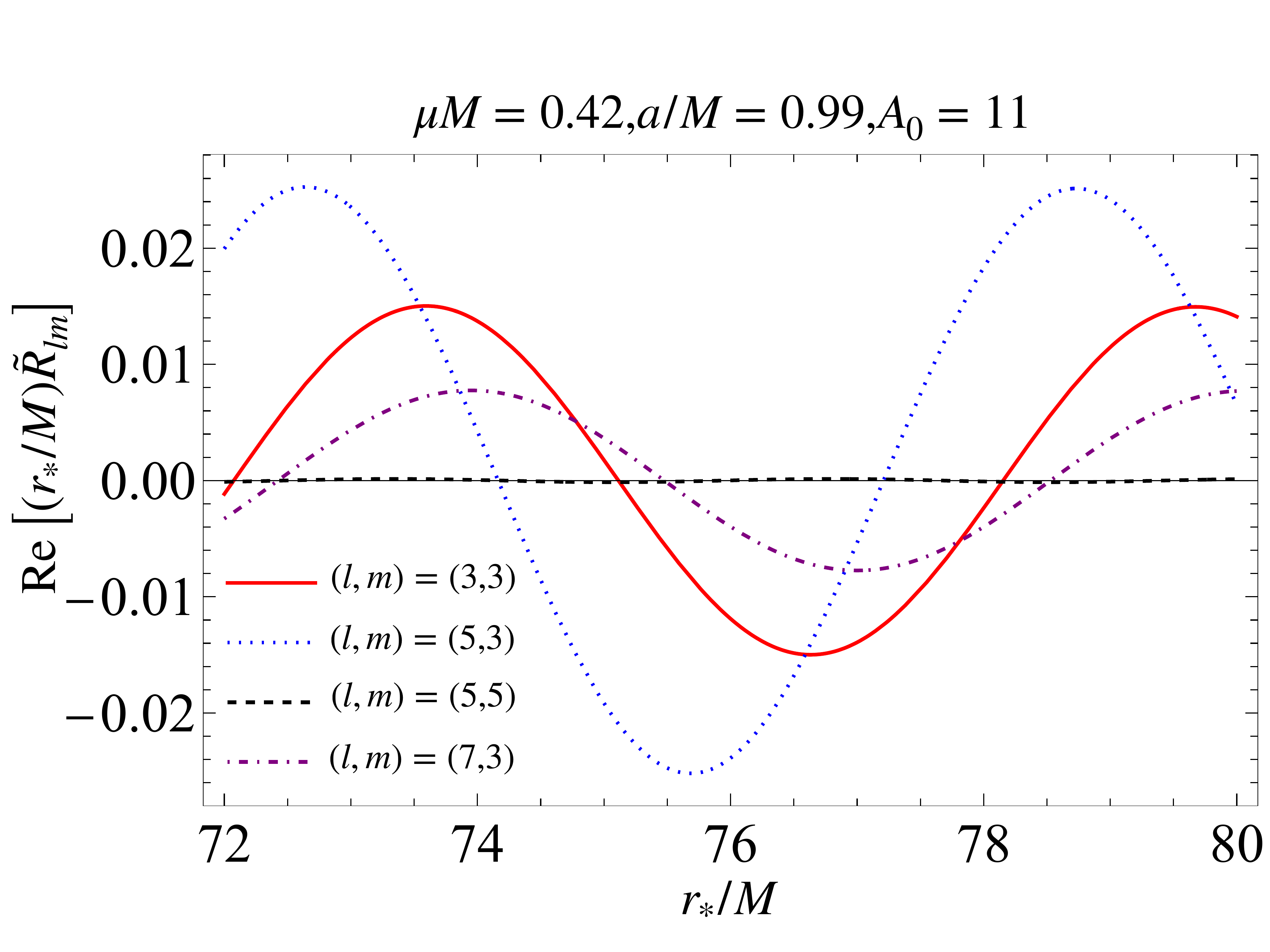}
	\caption{Each curve shows the real part of  $(r_*/M) \tilde{R}_{lm}$ near infinity. The red solid curve, blue dotted curve, black dashed curve, and  purple dashed dotted curve correspond to $(l,m) = (3,3),(5,3),(5,5),$ and $(7,3)$ modes, respectively.}
	\label{fig:fluxinf}
\end{figure}

Figure \ref{fig:energyandfluxhighl}  shows how the total flux $F_{\rm tot}$ depend on the energy $E$. We confirm that the total flux differs by a factor of $\sim 1.3$ when amplitude is large. This is because the flux to infinity becomes larger due to the additional radiative mode $(l,m) = (7,3)$. 

\begin{figure}[t]
	\centering
	\includegraphics[keepaspectratio,scale=0.3]{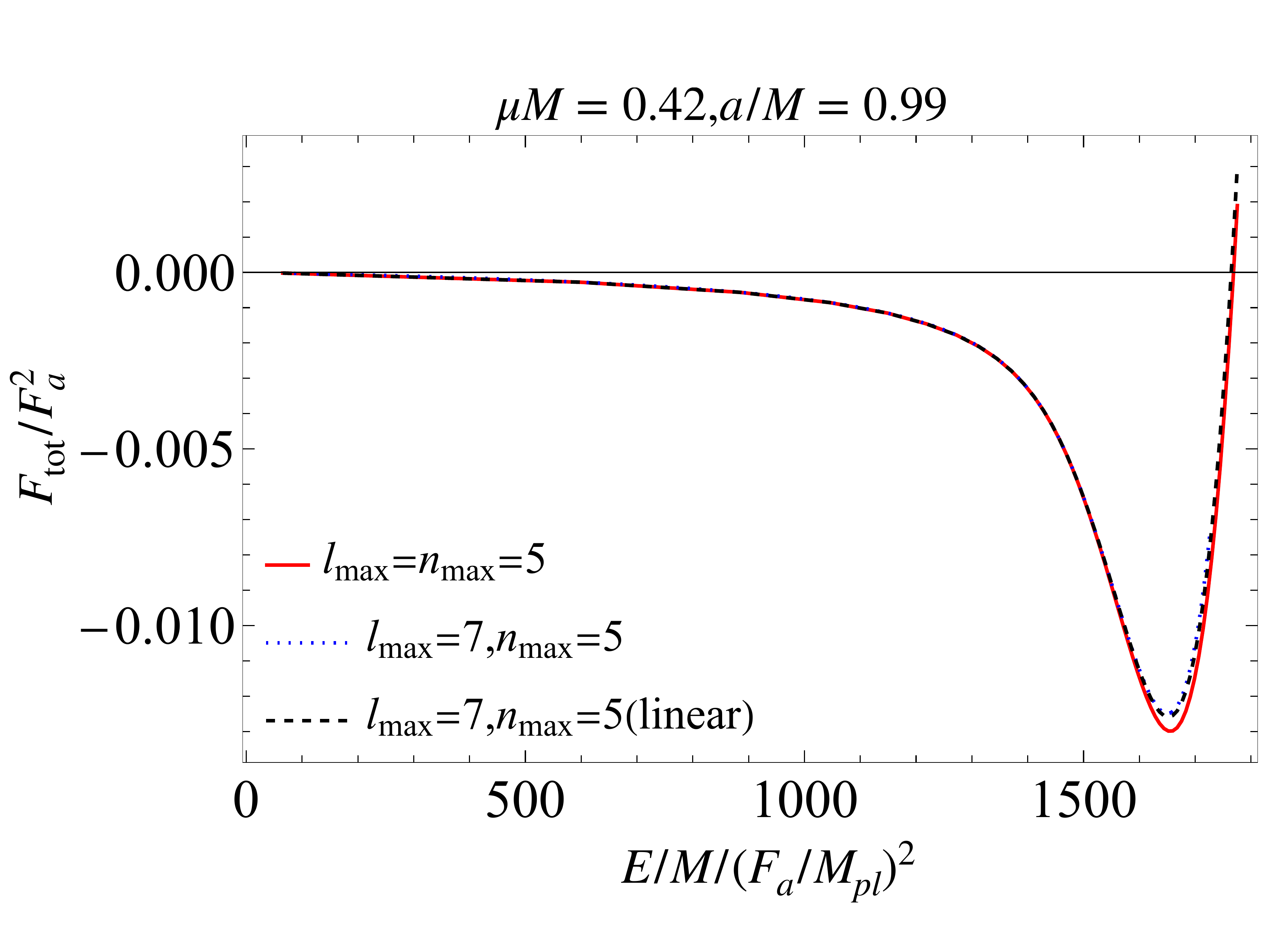}
	\caption{Dependence of the total flux $F_{\rm tot}(A_0)$ on the energy $E$. The red solid and blue dashed lines corresponds to the respective lines in right panel of Fig. \ref{fig:energyandfluxhighl}. Newly added black line corresponds to the total flux calculated by solving the linearized equation from the configuration neglecting the higher $l,m$ modes. }
	\label{fig:energyandfluxhighl_lin}
\end{figure}

To see the contribution of each mode on the flux to infinity, we show the behavior of mode functions at $A_0 = 11$ with $m \ge 3$ near infinity in Fig. \ref{fig:fluxinf}. From the figure, we observe that the $(l,m) = (5,3)$ mode gives the largest contribution, contrary to the naively expected dominance of the $(l,m) = (3,3)$ mode. 
The fact that the higher $l$ mode gives dominant contribution is very similar to the gravitational radiation from the axion cloud \cite{Yoshino:2013ofa}. The next dominant mode is the $(l,m) = (3,3)$ mode, and $(7,3)$ mode is further suppressed but not completely negligible to determine the saturation configuration. 
However, the configuration of the main body of the cloud is not affected much by the inclusion of $(l,m) = (7,3)$ mode, as shown in Fig. \ref{fig:modefnc}. Therefore the energy flux through $(l,m) = (7,3)$ can be computed from the linearized equation from the configuration obtained by neglecting the $(l,m) = (7,3)$ mode, as shown in Fig. \ref{fig:energyandfluxhighl_lin}. Moreover, we confirm that the $(l,m)=(5,5)$ mode gives much smaller contribution than the $m=3$ modes. Thus, inclusion of higher $m$ modes does not change our results.

\bibliographystyle{ptephy}
\bibliography{axionref}
\end{document}